\documentclass[journal,comsoc]{IEEEtran}

\makeatletter
\long\def\@makecaption#1#2{\ifx\@captype\@IEEEtablestring%
\footnotesize\begin{center}{\normalfont\footnotesize #1}\\
{\normalfont\footnotesize\scshape #2}\end{center}%
\@IEEEtablecaptionsepspace
\else
\@IEEEfigurecaptionsepspace
\setbox\@tempboxa\hbox{\normalfont\footnotesize {#1.}~~ #2}%
\ifdim \wd\@tempboxa >\hsize%
\setbox\@tempboxa\hbox{\normalfont\footnotesize {#1.}~~ }%
\parbox[t]{\hsize}{\normalfont\footnotesize \noindent\unhbox\@tempboxa#2}%
\else
\hbox to\hsize{\normalfont\footnotesize\hfil\box\@tempboxa\hfil}\fi\fi}
\makeatother

\ifCLASSINFOpdf
  \usepackage[pdftex]{graphicx}
  \usepackage{subfig}
\else
\fi

\usepackage{amsmath} 
\usepackage{algorithm}
\usepackage{algpseudocode}

\usepackage{amsthm}

\usepackage{color}

\begin{document}

\title{Multi-Drone 3D Trajectory Planning and Scheduling in Drone Assisted Radio Access Networks}

\author{Weisen~Shi,
        Junling~Li,
        Nan~Cheng,
        Feng~Lyu,
        Shan~Zhang,
        Haibo~Zhou,
        and~Xuemin~(Sherman)~Shen,~\IEEEmembership{Fellow,~IEEE}
\thanks{W. Shi, J. Li, N. Cheng (corresponding author), F. Lyu, and X. Shen are with the Department of Electrical and Computer Engineering, University of Waterloo, 200 University Avenue West, Waterloo, Ontario, Canada, N2L 3G1, (email: \{w46shi, j742li, n5cheng, f2lyu, sshen\}@uwaterloo.ca).}
\thanks{S. Zhang is with the School of Computer Science and Engineering, Beihang University, Beijing, China, 100191, (email: zhangshan2007@gmail.com).}
\thanks{H. Zhou is with the School of Electronic Science and Engineering, Nanjing University, Nanjing, China, 210023, (email: haibozhou@nju.edu.cn).}
\vspace{-0.5cm}
}

\maketitle

\begin{abstract}
Drone base station (DBS) is a promising technique to extend wireless connections for uncovered users of terrestrial radio access networks (RAN). 
To improve user fairness and network performance, in this paper, we design 3D trajectories of multiple DBSs in the drone assisted radio access networks (DA-RAN) where DBSs fly over associated areas of interests (AoIs) and relay communications between the base station (BS) and users in AoIs. 
We formulate the multi-DBS 3D trajectory planning and scheduling as a mixed integer non-linear programming (MINLP) problem with the objective of minimizing the average DBS-to-user (D2U) pathloss.
The 3D trajectory variations in both horizontal and vertical directions, as well as the state-of-the-art DBS-related channel models are considered in the formulation. 
To address the non-convexity and NP-hardness of the MINLP problem, we first decouple it into multiple integer linear programming (ILP) and quasi-convex sub-problems in which AoI association, D2U communication scheduling, horizontal trajectories and flying heights of DBSs are respectively optimized.
Then, we design a multi-DBS 3D trajectory planning and scheduling algorithm to solve the sub-problems iteratively based on the block coordinate descent (BCD) method.
A k-means-based initial trajectory generation and a search-based start slot scheduling are considered in the proposed algorithm to improve trajectory design performance and ensure inter-DBS distance constraint, respectively.
Extensive simulations are conducted to investigate the impacts of DBS quantity, horizontal speed and initial trajectory on the trajectory planning results.
Compared with the static DBS deployment, the proposed trajectory planning can achieve 10-15 dB reduction on average D2U pathloss, and reduce the D2U pathloss standard deviation by 68\%, which indicate the improvements of network performance and user fairness.
\end{abstract}

\IEEEpeerreviewmaketitle

\section{Introduction}
The future radio access networks (RAN) is expected to provide ubiquitous connectivity for any users or devices at any time with diversified service requirements \cite{bor2016new}. 
However, coverage holes (CH) of terrestrial RAN prevail in both urban and rural scenarios due to the lack of infrastructures or deeply blocking by obstacles \cite{zhang2018air}.
Specifically, as the terrestrial RAN are statically fixed in certain geographical locations, it is normally hard to ensure the quality of service (QoS) of users that are uneven and dynamically distributed in both spatial and temporal domain \cite{ye2018endmagazine}.
One solution to address those CHs is to deploy massive small cells, which, however, is inefficient and costly for RAN operators \cite{mozaffari2018tutorial}.
To overcome the coverage and flexibility challenges faced by current RAN, the emerging drone, \emph{a.k.a.} unmanned aerial vehicle (UAV), communication technology is proposed as a promising solution. 

Equipped with specific wireless transceivers, drones can communicate with both terrestrial users and cellular base stations (BSs) using WiFi \cite{fadlullah2016dynamic} or LTE \cite{al2017modeling} technologies. 
By integrating drone communication with terrestrial RAN, the drone assisted radio access networks (DA-RAN), in which drones perform as drone base stations (DBSs) to relay data between users in areas of interests (AoIs) and the associated terrestrial BS\footnote{In this paper, the abbreviation BS means the terrestrial base station and DBS is referred to as drone base station.}, has been proposed and verified by field experiments \cite{Dhekne2016}.
In DA-RAN, AoI includes both the CHs and the traffic dense spots (TDBs) where the allocated RAN spectrum resources are temporarily inadequate, e.g. congested road, concerts and sports events, etc.
Compared with the terrestrial RAN, DA-RAN advances in following four aspects:
1) The line-of-sight (LoS) probability for the DBS-to-ground (D2G) wireless link is higher than the terrestrial BS-to-user wireless link \cite{mozaffari2018tutorial}.
Experiments indicate that LoS links probability is the dominating factor to increase network performance \cite{lyu2018mobility};
2) DBSs can be dynamically deployed and dispatched to different controllers/users with respect to the spatial and temporal traffic variations \cite{xue2018device}; 
3) unlike connected vehicles whose mobility is controlled by drivers or autonomous driving controller, the trajectories of DBSs can be fully controlled by system providers, which empowers DBSs with the dynamic deployment feature \cite{cheng2019space} \cite{takaishi2018virtual}; 
4) DBS are capable of executing computing tasks by equipping with CPU or caching modules \cite{Jeong2018mobile} \cite{chen2017caching}. 
However, it is challenging to fully utilize the potential of DBSs due to the following two reasons.
First, the 3D mobility of DBS poses great complexity on the DBS spatial placement, especially in multi-DBS scenarios \cite{motlagh2016low}. 
Second, specific channel models are required to highlight the unique features of DBS-to-user (D2U) and DBS-to-BS (D2B) channels\cite{liu2018spaceairground}.

Several studies optimizing the multi-DBS spatial placements to support terrestrial users emerges in recent year, which can be divided into two categories, i.e., static DBS deployment and DBS trajectory planning. 
The static DBS deployment research focus on optimizing the hovering positions of DBSs to maximize terrestrial users QoS.
However, the static deployment fails to guarantee the fairness for users, in which the users located at the edge of the DBS's radio coverage suffer relatively higher pathloss compared with the users located at the center of the DBS's coverage. 
In addition, most existing DBS deployment works focus on optimizing the D2U communication, while ignore or idealize the D2B link quality constraints.

To promote the fairness for all users and maintain low deployment cost, some researchers further propose the DBS trajectory planning approach that allows DBSs fly over and serve AoIs periodically according to designed trajectories.
The purpose of DBS trajectory planning is optimizing AoIs association and trajectory for each DBS to maximize user QoS \cite{wu2017joint} \cite{wu2018common}.
However, three issues remain unsolved in current works.
First, to reduce the complexity of optimization problems, most existing works assume that all DBSs fly at a pre-defined constant height, which shrinks the 3D trajectory planning into a 2D horizontal trajectory planning, and fails to realize the performance improvements by adjusting DBS flying heights. 
Second, the commonly used assumption of Friis free space propagation model cannot reflect the unique D2G channel features.
Third, the D2B link quality constraint is also omitted by most DBS trajectory planning works.



To address those issues, in this paper, we investigate the 3D trajectory planning and scheduling for multiple DBSs in the DA-RAN.
Considering the state-of-the-art D2U \cite{al2014optimal} and D2B \cite{al2017modeling} channel models, and constraints of D2B link qualities, the multi-DBS 3D trajectory planning problem is formulated as a mixed integer non-linear programming (MINLP) problem which aims at minimizing the average D2U pathloss for all users within one trajectory period.
By decoupling the MINLP problem into multiple quasi-convex or integer linear programming (ILP) sub-problems, we can separately optimize the AoI association, D2U communication scheduling, DBS horizontal trajectories and flying heights in each sub-problem, respectively.
In essence, we adopt the block coordinate descent (BCD) mechanism to devise a multi-DBS 3D trajectory planning and scheduling algorithm, in which the sub-problems are iteratively optimized and converge to the optima. 
The main contributions of this work are listed as follows:
\begin{itemize}
\item We investigate the 3D trajectory planning of multiple DBSs in which both the flying heights and horizontal trajectories of DBS are optimized together instead of optimizing horizontal trajectories on a 2D plane.
As far as we know, this is the first work considering the real 3D trajectory in which the flying height of any DBS can be adjusted at different slots on its trajectory.
\item To make the system model more practical, we employ the state-of-the-art D2U and D2B pathloss models rather than the traditional pathloss models (e.g., Friis equation) in the system model. 
We formulate the multi-DBS trajectory planning problem, which turn to be an MINLP, and decouple it into multiple sub-problems to resolve the non-convexity.
A protect distance constraint between any two DBSs at every time slots is considered in the problem formulation to suppress the physical collision and mutual interference of DBSs. 
Instead of modifying the 3D trajectories, we ensure the protect distance constraint by scheduling the start slot of each trajectory to avoid introducing non-convex constraints in trajectory-related sub-problems.
\item A BCD based algorithm is proposed to separately optimize AoI association, D2U communication scheduling, horizontal trajectories and flying heights of DBSs in different sub-problems, respectively.
Besides, a \emph{k}-means-based scheme is devised to generate the DBS initial trajectories for further improvements on performance.
\item We conduct extensive simulations and results demonstrate that the proposed 3D trajectory planning and scheduling algorithm can reduce the average D2U pathloss by $15-20~\mathrm{dB}$, and lower the D2U pathloss standard deviation by $68\%$, in comparison with the static DBS deployment algorithm based on particle swarm optimization (PSO).
\end{itemize}

The remainder of this paper is organized as follows. 
The literature review is conducted in Section II.
In Section III the system model for DBS trajectory planning and scheduling in DA-RAN is introduced. 
Then the multi-DBS 3D trajectory planning and scheduling problem is formulated in Section IV. 
In Section V the MINLP problem in Section IV is decoupled into sub-problems and the BCD based algorithm is proposed to solve it.
Simulation and numerical results are carried out in Section VI, and the conclusion is given in Section VII.

\section{Related Works}
Promoted by the advancements in the flying control and communication technologies, both industry and academia are devoting many efforts to exploit the full potential of DA-RAN \cite{mozaffari2018tutorial}. 
As the foundation for drone communication and DA-RAN research, Al-Hourani \emph{et al}. built the D2U pathloss model for DBS according to abundant field test data in various scenarios \cite{al2014optimal}. 
A close-form expression of D2U pathloss model suiting different scenarios is proposed in which the probabilities of both LoS and NLoS D2U links are considered.
As the extension work, they further formulated the pathloss model for D2B communication in suburban scenario \cite{al2017modeling} where the D2B links are dominated by LoS links.
Leveraging the pathloss model in \cite{al2014optimal} and \cite{al2017modeling}, various studies have emerged in both static DBS deployment and DBS trajectory planning. 

\subsection{Static DBS Deployment}

In most static DBS deployment works, the terrestrial user QoS or network performance is improved through optimizing the hovering position of single or multiple DBSs.
For instance, through a clustering based approach, Mozaffari \emph{et al}. designed the optimal locations of DBSs that maximize the information collection gain from terrestrial IoT devices \cite{mozaffari2017mobile}. 
In \cite{zhang2017spectrum}, Zhang \emph{et al}. optimized the DBS density in DBS network to maximize the network throughput while satisfying the efficiency requirements of the cellular network.
Zhou \emph{et al}. studied the downlink coverage features of DBS using Nakagami-m fading models, and calculated the optimal height and density of multiple DBSs to achieve maximal coverage probability \cite{zhou2018coverage}.
Although various works have investigated the static DBS deployments in different scenarios with different methods, the D2B link quality constraint is simplified or ignored by most works. 
In the works considering the D2B links, the D2B channel models are either as same as the D2U pathloss model \cite{shah2017distributed} or traditional terrestrial channel models \cite{fouda2018uav}.
In this paper, we further implement the specific D2B channel model derived in \cite{al2017modeling} to highlight the D2B channel features.

\subsection{DBS Trajectory Planning}
In \cite{li2016energy}, Li \emph{et al}. proposed an cooperative relaying scheme in which multiple DBSs relay data from terrestrial sensors to the BS using time division multiple access (TDMA). 
As a pioneer work, the trajectories of all DBSs are assumed to be pre-optimized for simplicity.
Mozaffari \emph{et al} studied both the static and mobile DBS-enabled wireless networks underlaid with a device-to-device communication network \cite{mozaffari2016unmanned}.
Though the trajectory optimization is considered in this work, the D2U communications are only permitted at pre-defined stop points, which fails to exploit the impact of DBS mobility feature on the network performance.
Motivated by \cite{mozaffari2016unmanned}, Zeng \emph{et al}. proposed a general framework for joint trajectory and communication optimization in D2U point-to-point communication scenario \cite{zeng2017energy}. 
For DBS-enabled multi-user networks, Lyu \emph{et al}. proposed a cyclical multiple access scheme in \cite{lyu2016cyclical}, where the DBS forms a cyclical trajectory and periodically serves each terrestrial users using TDMA. 
Wu \emph{et al}. formulated the DBS trajectory planning as a mixed integer non-convex optimization problem in which the user association, DBSs’ trajectories planning and DBS transmitting power control schemes are jointly optimized \cite{wu2017joint}.
Considering the delay constraints, Wu \emph{et al}. further studied a DBS-enabled OFDMA network where a single DBS is dispatched to serve a group of delay-sensitive users on the ground \cite{wu2018common}. 
The DBS trajectory planning and OFDMA resource allocation are jointly optimized through an iterative parameter-assisted BCD method.
As the pioneer works of trajectory planning, \cite{wu2017joint} \cite{wu2018common} set the foundation models for DBS trajectory planning.
To dynamically deploy multiple DBSs while maintaining the connectivity among them, Zhao \emph{et al}. proposed both the centralized and distributed DBS motion control algorithms for scenarios with or without global information of users \cite{zhao2018deployment}.
However, in those works the flying height of all DBS are treated as a pre-defined constant, and most of them idealize the D2B link quality constraints.
In this paper, not only the D2B link quality constraint is introduced in problem formulation and optimization, but also the flying heights of each DBS at every slots are jointly optimized with the horizontal trajectory.
\begin{figure}[tpb]
  \centering
  \subfloat[DBS photograph \cite{EEdrone}]{\includegraphics[width=0.37\textwidth]{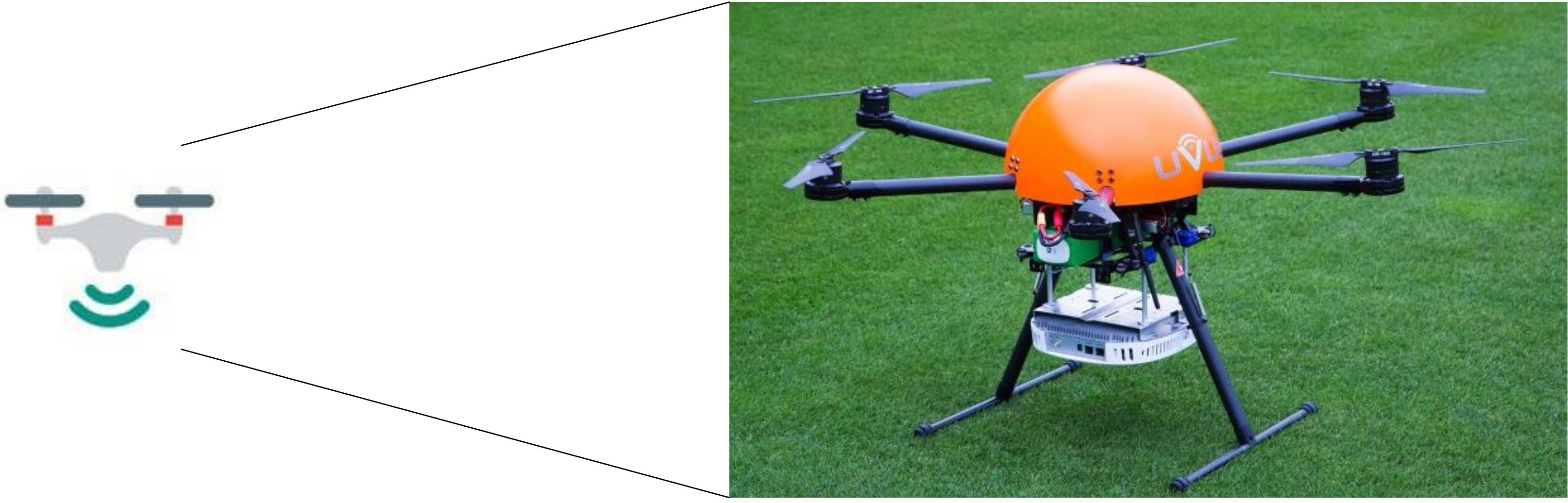}}\\
  \subfloat[System model]{\includegraphics[width=0.37\textwidth]{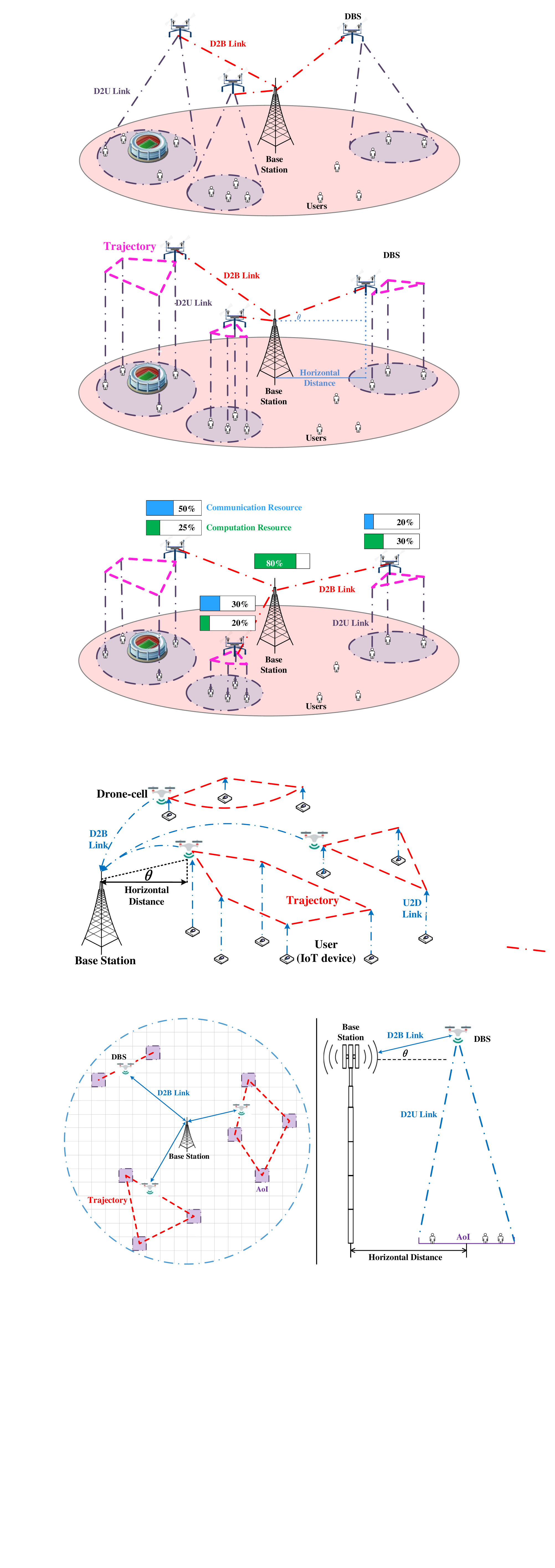}}
  \caption{Multi-DBS 3D trajectory planning and scheduling.}
  \label{AoIRAN}
  \vspace{-0.5cm}
\end{figure}

\section{System Model}
In this section, we introduce the DA-RAN scenario, D2U and D2B channel models, as well as the DBS trajectory model used for further analyses and problem formulation.
\subsection{Drone Assisted Radio Access Networks}
Fig. \ref{AoIRAN} shows the system model and the DBS photograph for the multi-DBS 3D trajectory planning and scheduling.
The DBS is supported by the quad-copter or multi-rotor drones, which is regarded as low altitude platform (LAP) with the low flying height (below 300m) \cite{motlagh2016low}, limited communication coverage (less than 3km) \cite{motlagh2016low} and static hovering capability.
Based on the DA-RAN architecture, we investigate the scenario in which multiple DBSs are controlled by a single BS to support users in AoI through the state-of-the-art wireless relay techniques \cite{liMaoGQ2017relay} \cite{liNoMGQ2016relay}.
The radio coverage area of the BS ${|\mathcal{S}|}_{\mathrm{bs}}$ is a circle with radius $r_{\mathrm{BS}}$, and is divided into a mesh consisting of multiple grids on X-Y plane. 
The side-length of each grid is denoted as ${L_{\mathrm{aoi}}}$.
Without loss of generality, the average D2U pathloss for any users in one grid can be treated as equal since ${L_{\mathrm{aoi}}}$ is far smaller than $r_{\mathrm{BS}}$.
Assuming that users are uniformly distributed over ${|\mathcal{S}|}_{\mathrm{bs}}$, and each grid can be chosen as AoI with same probability.
Therefore, the user association is equal to AoI association in this work.
Both users in AoIs and DBSs are considered as identical devices with identical transmit power and uplink/downlink bandwidth.
Considering the fact that AoIs change their distribution in a relatively low frequency, the dynamic distribution of AoIs can be treated as a quasi-static scenario between successive trajectory planning.
Based on the current snapshot of AoIs distribution, the BS running trajectory planning and scheduling algorithm to calculate optimal trajectories for all DBSs, and update them to the DBSs via D2B links.
When the BS senses significant changes of AoIs distribution, re-planning process is triggered to design new trajectories for DBSs, otherwise, the trajectory planning result keeps constant.
The set of AoIs to be served and the set of DBSs to be deployed are denoted as $\mathcal{U}$ and $\mathcal{D}$, respectively.
Their cardinalities, $|\mathcal{U}|$ and $|\mathcal{D}|$, represent the number of AoIs and DBSs, respectively.

\subsection{D2U and D2B Channel Models}
Based on the state-of-the-art Drone-to-Ground (D2G) channel research \cite{al2014optimal} \cite{al2017modeling}, both the D2U and D2B links are modeled in our work, respectively.
For D2U links, the LoS probability is calculated as \cite{al2014optimal}
\begin{equation}
{P_{\mathrm{LoS}}(r_{\mathrm{DU}},h)} = \frac{1}{1 + a\exp(-b(\arctan(\frac{h}{r_{\mathrm{DU}}}) - a))}
\label{P_LoS}
\end{equation}
where $r_{\mathrm{DU}}$ is the horizontal distance between DBS and the AoI, $h$ represents the flying heights of the DBS. $a$ and $b$ are environment-based constant values. 
The average D2U pathloss can be derived based on (\ref{P_LoS})
\cite{al2014optimal}:
\begin{equation}
\begin{aligned}
{PL(r_{\mathrm{DU}},h)} &= 20\log(\frac{4 \pi f_{c} \sqrt{h^2+r_{\mathrm{DU}}^2}}{c})\\
&+ P_{\mathrm{LoS}}(r_{\mathrm{DU}},h)\eta_{\mathrm{LoS}} + (1-P_{\mathrm{LoS}}(r_{\mathrm{DU}},h))\eta_{\mathrm{NLoS}}
\end{aligned}
\label{pathlossEq}
\end{equation}
where $f_{c}$ (Hz) and $c$ (m/s) are carrier frequency and speed of light, respectively. 
$\eta_{\mathrm{LoS}}$ and $\eta_{NLoS}$ are additional losses for LoS and NLoS links obtained through field test data, which involves the impacts of shadowing components. $a$, $b$, $\eta_{\mathrm{LoS}}$ and $\eta_{NLoS}$ are all environment-based parameters.

D2B links are designed to provide high-reliability data transmission between DBSs and their corresponding BS.
The average D2B pathloss is calculated as follow by implementing the D2B channel model in \cite{al2017modeling}:
\begin{equation}
\begin{aligned}
{PL(r_{\mathrm{DB}},\theta)} &= 10\alpha\log(r_{\mathrm{DB}}) + A(\theta - \theta_{0})\mathrm{e}^{(\frac{\theta_{0}-\theta}{B})}+\eta_{0}
\end{aligned}
\label{pathlossEqD2B}
\end{equation}
where $r_{\mathrm{DB}}$ and $\theta$ denote the horizontal distance and the vertical angle between the DBS and the BS antenna, respectively.
$\alpha$, $A$, $\theta_{0}$, $B$, and $\eta_{0}$ are the terrestrial pathloss exponent, excess pathloss scalar, angle offset, angle scalar, and excess pathloss offset, respectively. 
All of them are environment-based parameters and involving impacts of shadowing components.
Except $r_{\mathrm{DB}}$ and $\theta$, all other parameters in (\ref{pathlossEqD2B}) are environment-based constants.
Since the D2B channel model in \cite{al2017modeling} use $850~\mathrm{MHz}$ LTE bands, (\ref{pathlossEqD2B}) contains no parameter representing carrier frequency.

Both the D2U \cite{al2014optimal} and D2B \cite{al2017modeling} pathloss models are large-scale pathloss models.
For the small scale fading and multipath effects, currently there is no specific model for the drone-to-ground communication links. 
Moreover, since the objective of our multi-DBS trajectory planning is to minimize the mean D2U pathloss of the system, the small-scale pathloss can be average out at zero or a constant offset during the analysis.
Therefore, based on the assumptions in D2U and D2B channel models, in this work we do not focus on the small scale fading and multipath effects in D2U and D2B links.

\subsection{DBS Trajectory Model}
For an arbitrary DBS $d \in \mathcal{D}$, we design its trajectory such that the DBS serves the associated AoI set $\mathcal{A}_d \subseteq \mathcal{U}$ periodically.
Within one period $T$, $d$ flies over all its associated AoI and serves them sequentially according to the scheduling result.
Since the continuous time can introduce infinite number of position variables to describe the DBS trajectory, we discrete the period $T$ into $N$ equal-time slots to simplify the formulation. 
The length of each slot $\delta_{t} = \frac{T}{N}$ can be set as small as possible to approximate the continuous optimal trajectory.
Based on this model, the trajectory of DBS $d$ within each $T$ can be modeled as a $N$-length sequence composed by three-dimensional vectors:
\begin{equation}
\begin{aligned}
\mathbf{W}_d[n] = [x_d[n], y_d[n], h_d[n]], \quad n = 1,...,N
\end{aligned}
\label{trajectoryModel}
\end{equation}
where $x_d[n]$, $y_d[n]$ and $h_d[n]$ denote the 3D coordinates of DBS $d$ at slot $n$. 
DBS $d$ is considered to follow the same trajectory $\mathbf{W}_d[n]$ over consecutive periods until the re-planning process is triggered. 
For multiple DBSs working simultaneously, they share the same trajectory period length $T$ to simplify the trajectory planning and scheduling. 

Several trajectory constraints are considered in our work: 
1) Each DBS needs to return to its initial location by the end of each period $T$, which implies that the trajectory of each DBS is a closed curve in 3D space. 
2) Within any slot $n$, the horizontal and vertical shifts of any DBS cannot exceed the maximal horizontal distance $V_\mathrm{max}\delta_{t}$, and the maximal height difference $H_\mathrm{max}\delta_{t}$, respectively. $V_\mathrm{max}$ and $H_\mathrm{max}$ are maximum allowed horizontal and vertical speeds. 
3) For any slot $n$, the 3D distance between any two DBSs cannot be smaller than a pre-defined protect distance $Z_\mathrm{min}$, which prevents the physical collision, disturbance and mutual interference among DBSs.
Since calculating interference from non-associated DBSs to any AoI based on D2U pathloss model is highly complex and makes the trajectory planning problem unsolvable, in this work, we assume that the mutual interference (to AoIs) among DBSs can be effectively avoided by ensuring the protect distance constraint. 

\subsection{AoI Association and D2U Communication Scheduling}
In this work, the DBS-AoI association is denoted by the binary variable $a_{d,u}$.
$a_{d,u} = 1$ when AoI $u \in \mathcal{U}$ is associated to DBS $d \in \mathcal{D}$, and otherwise $a_{d,u} = 0$.
The U2D communication scheduling is denoted by the binary variable $k_{d,u}[n]$ for $\forall d \in \mathcal{D}, u \in \mathcal{U}, n \in \mathcal{N}$.
If AoI $u$ is severed by DBS $d$ in slot $n$, $k_{d,u}[n]$ is set as $1$; otherwise, $k_{d,u}[n] = 0$.
For each DBS with pre-defined trajectory planning and AoI association results, a D2U communication scheduling scheme is designed to allocate each slot to the corresponding AoI, and guarantee the fairness among all associated AoIs.
Several constraints are considered in the AoI association and scheduling model:
1) One DBS can serve maximal ${|\mathcal{A}_d|}_\mathrm{max}$ number of AoIs.
2) In any slot $n$, one DBS $d$ can serve at most one AoI $u \in \mathcal{A}_d$; in all slots, one AoI $u$ can only be associated to one DBS.
3) For any given $T$ and $\mathcal{A}_d$, the total $N$ slots are uniformly scheduled to each $u \in \mathcal{A}_d$ to ensure fairness.
4) The slot amount scheduled to every AoIs cannot be smaller than a pre-defined threshold $S_\mathrm{min}$, which indicates the minimal user service time constraint.
5) To prevent the overloads and delay caused by frequent switching between associated AoIs, all slots scheduled to one $u$ within $T$ have to be consecutive.

\section{Problem Formulation}
In this section we formulate the multi-DBS 3D trajectory planning problem based on the aforementioned system model.

According to  (\ref{trajectoryModel}), the the 3D distance from the DBS $d$ to AoI $u$ in time slot $n$ can be expressed as
\begin{equation}
\begin{aligned}
m_{d,u}[n] & = \sqrt{h_d[n]^2 + {\| \mathbf{l}_d[n] - \mathbf{l}_u \|}^2} \\
           & = \sqrt{h_d[n]^2 + r_{d,u}[n]^2}
\end{aligned}
\label{trajectoryDist}
\end{equation}
where $\mathbf{l}_u = [x_u, y_u]$ is the 3D coordinate of AoI $u$.
$\mathbf{l}_d[n]$ is $d$'s 2D projection on X-Y plane $\mathbf{l}_d[n] = [x_d[n], y_d[n]]$. 
$r_{d,u}[n]$ denotes the horizontal distance between DBS $d$ and AoI $u$.
Without loss of generality, we set the BS at the original point of coordinate system.
By substituting $m_{d,u}[n]$ for the D2U and D2B distances in (\ref{pathlossEq}) and (\ref{pathlossEqD2B}), we can calculate the D2U pathloss between DBS $d$ and AoI $u$ in slot $n$ 
\begin{equation}
\begin{aligned}
P_{d,u}[n] & = 20\log(\frac{4 \pi f_{c} m_{d,u}[n]}{c}) \\
           & + P_{\mathrm{LoS}}(r_{d,u}[n],h_d[n])\eta_{\mathrm{LoS}} \\
           & + (1-P_{\mathrm{LoS}}(r_{d,u}[n],h_d[n]))\eta_{\mathrm{NLoS}}
\end{aligned}
\label{D2UpathlossSub}
\end{equation}
as well as the D2B pathloss between the BS and DBS $d$ in slot $n$ 
\begin{equation}
\begin{aligned}
P_{d,B}[n] & = 10\alpha\log({\|\mathbf{l}_d[n]\|}) \\
           & + A(\theta_{d,B}[n] - \theta_{0})\mathrm{e}^{(\frac{\theta_{0}-\theta_{d,B}[n]}{B})}+\eta_{0}
\end{aligned}
\label{D2BpathlossSub}
\end{equation}
where $\theta_{d,B}[n] = \arctan(h_d[n]/{\|\mathbf{l}_d[n]\|})$ in degree.

Since all users and DBSs are identical devices with fixed transmission power and transmission bandwidth in each period, the achievable D2U data rate between DBS $d$ and AoI $u$ is negative correlated with the D2U pathloss.
Therefore, the aim of the multi-DBS 3D trajectory planning and scheduling problem is minimizing the average D2U pathloss of the network over one period $T$.

Define $\mathbf{A} = \{a_{d,u},~\forall d,u\}$, $\mathbf{K} = \{k_{d,u}[n],~\forall d,u,n\}$ and $\mathbf{W} = \{\mathbf{W}_d[n],~\forall d,n\}$, the trajectory planning and scheduling problem can be formulated as
\begin{subequations}
\begin{align}
& \min_{\mathbf{A},\mathbf{K},\mathbf{W}} 
\quad \frac{1}{N|\mathcal{U}|}\sum\limits_{u=1}^{|\mathcal{U}|}\sum\limits_{d=1}^{|\mathcal{D}|}{a_{d,u}}(\sum\limits_{n=1}^{N}k_{d,u}[n]P_{d,u}[n]) \tag{\ref{optimal}}\\
& s.t. \quad \sum\nolimits_{u=1}^{|\mathcal{U}|}a_{d,u} \le {|\mathcal{A}_d|}_\mathrm{max}, \quad \forall d, \label{optimal:a}\\
      & \quad \quad \sum\nolimits_{d=1}^{|\mathcal{D}|}a_{d,u} = 1, \quad \forall u, \label{optimal:b}\\
      & \quad \quad \sum\nolimits_{u=1}^{|\mathcal{U}|}k_{d,u}[n] = 1, \quad \forall d,n, \label{optimal:c}\\
      & \quad \quad \sum\nolimits_{d=1}^{|\mathcal{D}|}k_{d,u}[n] = 1, \quad \forall u,n, \label{optimal:d}\\
      & \quad \quad \sum\nolimits_{n=1}^{N}k_{d,u}[n] = \frac{N}{|\mathcal{A}_d|}, \quad \forall d,u, \label{optimal:e}\\
      & \quad \quad \sum\nolimits_{n=1}^{N}k_{d,u}[n] \ge S_\mathrm{min}, \quad \forall d,u, \label{optimal:f}\\
      & \quad \quad \sum\nolimits_{o}^{\frac{N}{|\mathcal{A}_d|}}k_{d,u}[{(n+o)}\bmod{N}] \le \frac{N}{|\mathcal{A}_d|}, \quad \forall d,u,n, \label{optimal:g}\\ 
      & \quad \quad a_{d,u}, k_{d,u}[n] \in \{0,1\}, \quad \forall d,u,n, \label{optimal:h}\\
      & \quad \quad \mathbf{W}_d[1] = \mathbf{W}_d[N+1], \quad \forall d, \label{optimal:i}\\
	  & \quad \quad {\| \mathbf{l}_d[n+1] - \mathbf{l}_d[n] \|} \le {V_\mathrm{max}\delta_{t}}, \quad \forall d,n, \label{optimal:j}\\
	  & \quad \quad {|h_d[n+1] - h_d[n]|} \le H_\mathrm{max}\delta_{t}, \quad \forall d,n, \label{optimal:k}\\
	  & \quad \quad {\|\mathbf{W}_i[n] - \mathbf{W}_j[n]\|} \ge Z_\mathrm{min}, \quad \forall n,i,j \ne i, \label{optimal:l}\\
	  & \quad \quad P_{d,B}[n] \le P_{\mathrm{DB}}, \quad \forall d,u,n. \label{optimal:m}
\end{align}
\label{optimal}
\end{subequations}
In  (\ref{optimal}), $|\mathcal{A}_d| = {N}/{\sum\nolimits_{u=1}^{|\mathcal{U}|}a_{d,u}}$ is the number of AoIs associated to DBS $d$. 
$\|\mathbf{W}_i[n] - \mathbf{W}_j[n]\|$ represents the 3D distance between DBS $i$ and $j$ at slot $n$.
$a\bmod{b}$ is the modulo operation between $a$ and $b$.
$P_{\mathrm{DB}}$ is the pathloss threshold for D2B communication.
(\ref{optimal:a})-(\ref{optimal:h}) are AoI association and D2U communication scheduling constraints, in which (\ref{optimal:a}) is constraint 1); (\ref{optimal:b})-(\ref{optimal:d}) represent the constraint 2);  (\ref{optimal:e}) and  (\ref{optimal:f}) corresponds to constraint 3) and 4), respectively;  (\ref{optimal:g}) indicates constraint 5).
(\ref{optimal:i})-(\ref{optimal:l}) correspond to DBS trajectory constraints 1), 2) and 3). 
(\ref{optimal:m}) is the D2B pathloss constraint.

Due to the quadratic and exponential terms in (\ref{optimal}) and constraints, as well as the binary variable $a_{d,u}$, $k_{d,u}[n]$, problem (\ref{optimal}) is a MINLP problem \cite{bor2016efficient}.
Besides, the optimization objective (\ref{optimal}) and constraints are non-convex for DBS trajectory $\mathbf{W}$, which is difficult to solve directly.

\section{Multi-DBS 3D Trajectory Planning and Scheduling Algorithm}
Although the objective and constraints in problem (\ref{optimal}) are non-convex or non-linear for the decision variables, the problem can still be transformed into solvable forms (e.g. quasi-convex or ILP) by setting parts of the decision variables as constants.
Then, the MINLP problem can be decoupled into multiple sub-problems which are solvable for parts of the decision variables.
Specifically, for the multi-DBS 3D trajectory planning and scheduling problem, we divide the decision variable set into four blocks (i.e. $\mathbf{A}$, $\mathbf{K}$, $\mathbf{L} = \{\mathbf{l}_d[n],~\forall d,n\}$ and $\mathbf{H} = \{h_d[n],~\forall d,n\}$), and propose multiple sub-problems in which all blocks or their sub-blocks are optimized, respectively.
However, the problem (\ref{optimal}) remains non-convex to DBS trajectory variable $\mathbf{W}$ even with given $\mathbf{A}$ and $\mathbf{K}$.
Therefore, we further divide $\mathbf{W}$ into two independent blocks, i.e. the horizontal DBS trajectory $\mathbf{L}$ and the DBS flying height $\mathbf{H}$.

\subsection{AoI Association Optimization}
Given the constant $\mathbf{K}$, $\mathbf{L}$ and $\mathbf{H}$, which indicate the pre-defined trajectories of multiple DBSs, the AoI association sub-problem can be written as an ILP problem:
\begin{equation}
\begin{aligned}
& \min_{\mathbf{A}}
\quad \frac{1}{N|\mathcal{U}|}\sum\limits_{u=1}^{|\mathcal{U}|}\sum\limits_{d=1}^{|\mathcal{D}|}{a_{d,u}}(\sum\limits_{n=1}^{N}k_{d,u}[n]P_{d,u}[n]) \\
& s.t. \quad (\ref{optimal:a}), (\ref{optimal:b}), \quad a_{d,u} \in \{0,1\} \quad \forall d,u.
\end{aligned}
\label{subpAssociation}
\end{equation}
Since exact $\mathbf{K}$ can only be determined with given $\mathcal{A}_d$, an initial D2U communication scheduling $\mathbf{K}_{0}$, in which $k_{d,u}[n] = 1,~\forall d,u,n$, is defined for the first AoI association optimization.
The branch and bound method supported by various solvers (e.g. Gurobi \cite{gurobi2018optimization}) can be used to solve problem (\ref{subpAssociation}) efficiently. 

\subsection{D2U Communication Scheduling Optimization}
Based on the optimized $\mathbf{A}$, as well as the constant $\mathbf{L}$ and $\mathbf{H}$, the D2U communication scheduling sub-problem is an ILP problem too:
\begin{equation}
\begin{aligned}
& \min_{\mathbf{K}} 
\quad \frac{1}{N|\mathcal{U}|}\sum\limits_{u=1}^{|\mathcal{U}|}\sum\limits_{d=1}^{|\mathcal{D}|}{a_{d,u}}(\sum\limits_{n=1}^{N}k_{d,u}[n]P_{d,u}[n]) \\
& s.t. \quad (\ref{optimal:a}),(\ref{optimal:b}),(\ref{optimal:c}),(\ref{optimal:d}),(\ref{optimal:e}),(\ref{optimal:f}),(\ref{optimal:g}),\\
& \quad \quad k_{d,u}[n] \in \{0,1\} \quad \forall d,u,n.
\end{aligned}
\label{subpScheduling}
\end{equation}
It is worth noting that constraint (\ref{optimal:e}) and (\ref{optimal:g}) turn to be linear constraint to $\mathbf{K}$ given constant $\mathbf{A}$.
Same as problem (\ref{subpAssociation}), problem (\ref{subpScheduling}) can be efficiently solved by the branch and bound method. 

\subsection{DBS Horizontal Trajectory Optimization}
The sub-problem to optimize $\mathbf{L}$ with constant $\mathbf{A}$, $\mathbf{K}$ and $\mathbf{H}$ can be expressed as
\begin{equation}
\begin{aligned}
& \min_{\mathbf{L}} 
\quad \frac{1}{N|\mathcal{U}|}\sum\limits_{u=1}^{|\mathcal{U}|}\sum\limits_{d=1}^{|\mathcal{D}|}{a_{d,u}}(\sum\limits_{n=1}^{N}k_{d,u}[n]P_{d,u}[n]) \\
& s.t. \quad (\ref{optimal:j}), (\ref{optimal:m}), \quad \mathbf{l}_d[1] = \mathbf{l}_d[N+1] \quad \forall d.
\end{aligned}
\label{subpHorTrajR}
\end{equation}
According to (\ref{D2UpathlossSub}), (\ref{subpHorTrajR}) is non-convex for $\mathbf{L}$.
Instead of jointly optimizing $\mathbf{L}$, we further divide the block into its element variable $\mathbf{l}_{d}[n]$ and revise (\ref{subpHorTrajR}) as
\begin{equation}
\begin{aligned}
& \min_{\mathbf{l}_{d}[n]} 
\quad \frac{1}{N|\mathcal{U}|}a_{d,u}[n]k_{d,u}[n]P_{d,u}[n] + \\
& \quad \quad \quad \frac{1}{N|\mathcal{U}|}\sum\limits_{u=1}^{|\mathcal{U}|}\sum\limits_{d=1}^{|\mathcal{D}|}{a_{d,u}}(\sum\limits_{\bar{n}=1,\bar{n} \neq n}^{N}k_{d,u}[\bar{n}]P_{d,u}[\bar{n}]) \\
& s.t. \quad (\ref{optimal:j}), (\ref{optimal:m}), \quad \mathbf{l}_d[1] = \mathbf{l}_d[N+1] \quad \forall d.
\end{aligned}
\label{subpHorTrajEleR}
\end{equation}
Keeping other $\mathbf{l}_{d}[\bar{n}],~\forall d,u,\bar{n} \neq n$ fixed, the second part of (\ref{subpHorTrajEleR}) turns to be constant.
With the given $\mathbf{H}$, we can prove that $P_{d,u}[n]$ is a quasi-convex and non-decreasing function to D2U horizontal distance $r_{d,u}[n], \forall d,u,n$.
Therefore, minimizing the objective function in (\ref{subpHorTrajEleR}) equals minimizing $r_{u,d}[n]^2 = {\| \mathbf{l}_d[n] - \mathbf{l}_u \|}^2$, which is a quadratic convex optimization problem for $\mathbf{l}_d[n]$:
\begin{equation}
\begin{aligned}
& \min_{\mathbf{l}_d[n]} 
\quad \| \mathbf{l}_d[n] - \mathbf{l}_u \|^2 \\
& s.t. \quad (\ref{optimal:j}), (\ref{optimal:m}), \quad \mathbf{l}_d[1] = \mathbf{l}_d[N+1], \quad \forall d.
\end{aligned}
\label{subpHorTrajEql}
\end{equation}
It is worth noting that the feasible region of $\mathbf{l}_d[n]$ constrained by (\ref{optimal:m}) can form a convex set in any X-Y plane by ignoring the working-zone burst close to the BS antenna \cite{shi2018multiple}.

\subsection{DBS Flying Height Optimization}
\begin{figure}[htbp]
  \centering
  \includegraphics[width=0.37\textwidth]{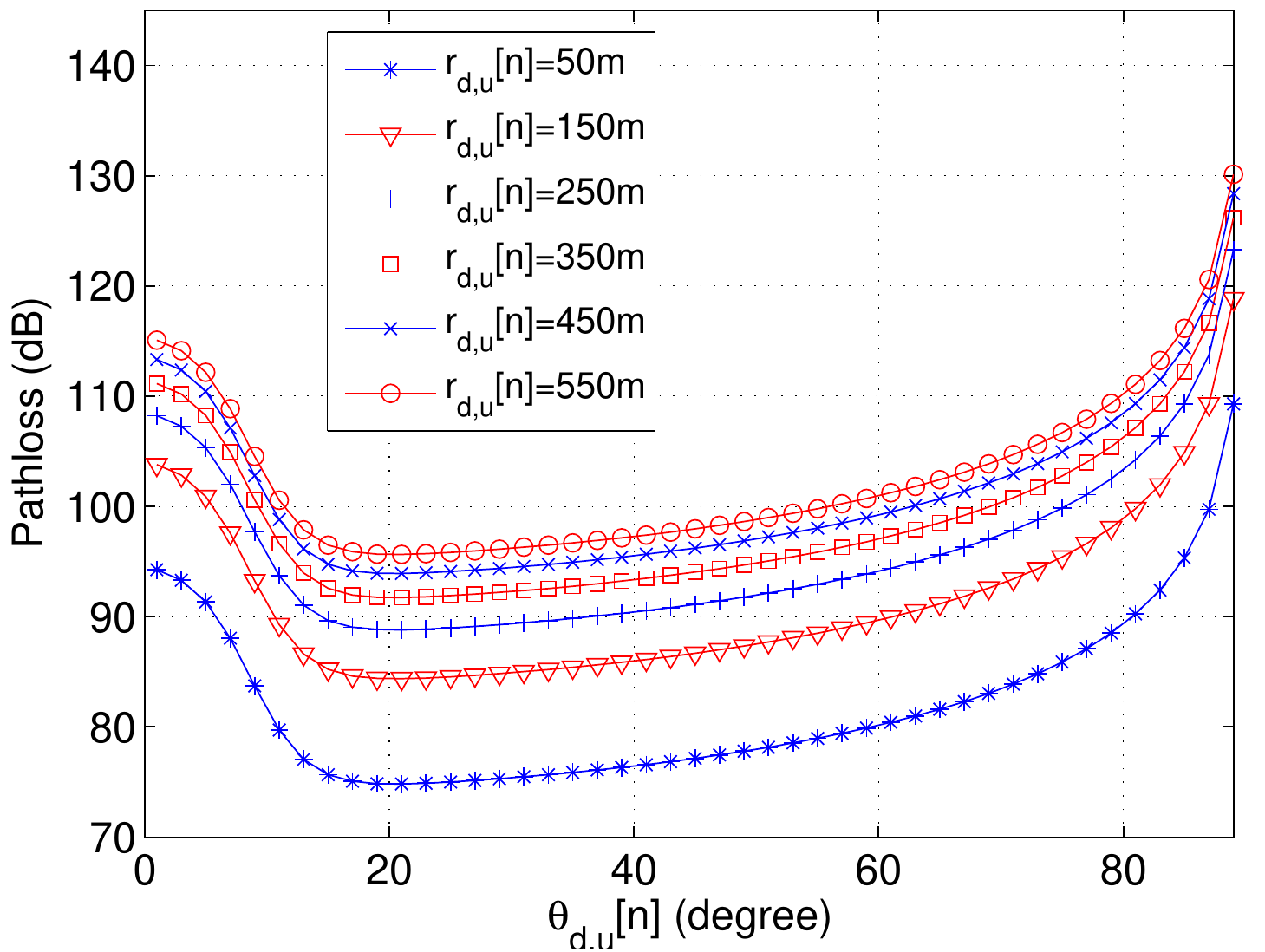}
  \caption{$P_{d,u}[n]$ versus $\theta_{d,u}[n]$.}
  \label{pathlossFigTheta}
  \vspace{-0.4cm}
\end{figure}

Similar to problem (\ref{subpHorTrajR}), the sub-problem optimizing $\mathbf{H}$ is also non-convex with given $\mathbf{A}$, $\mathbf{K}$ and $\mathbf{L}$. 
Further decoupling $\mathbf{H}$ into $h_{d}[n]~\forall d,n$, the sub-problem to optimize each $h_{u,d}[n]$ is 
\begin{equation}
\begin{aligned}
& \min_{h_{d}[n]} 
\quad \frac{1}{N|\mathcal{U}|}a_{d,u}[n]k_{d,u}[n]P_{d,u}[n] + \\
& \quad \quad \quad \frac{1}{N|\mathcal{U}|}\sum\limits_{u=1}^{|\mathcal{U}|}\sum\limits_{d=1}^{|\mathcal{D}|}{a_{d,u}}(\sum\limits_{\bar{n}=1,\bar{n} \neq n}^{N}k_{d,u}[\bar{n}]P_{d,u}[\bar{n}]) \\
& s.t. \quad (\ref{optimal:k}), (\ref{optimal:m}), \quad h_d[1] = h_d[N+1] \quad \forall d.
\end{aligned}
\label{subpHeightEleH}
\end{equation}
To solve (\ref{subpHeightEleH}), we first transform (\ref{D2UpathlossSub}) as the summation of one function of $r_{d,u}[n]$ and one function of $\theta_{d,u}[n] = \arctan(h_d[n]/r_{d,u}[n])$:
\begin{equation}
\begin{aligned}
&P_{d,u}[n] = 20\log(\frac{4{\pi}f_{c}}{c}r_{d,u}[n]) + \eta_{\mathrm{NLoS}} + F(\theta_{d,u}[n])
\end{aligned}
\label{D2UpathlossThetaSubR}
\end{equation}
where
\begin{equation}
\begin{aligned}
F(\theta_{d,u}[n]) & = 20\log(\sec(\theta_{d,u}[n]) \\
                   & + \frac{\eta_{\mathrm{LoS}} - \eta_{\mathrm{NLoS}}}{1 + a\exp(-b(\theta_{d,u}[n] - a))}.
\end{aligned}
\label{D2UpathlossThetaSubOnlyTheta}
\end{equation}
Since $r_{d,u}[n]$ is pre-defined with given $\mathbf{L}$, 
(\ref{D2UpathlossThetaSubR}) is constant and optimizing $h_{d}[n]$ equals optimizing $\theta_{d,u}[n]$ in (\ref{subpHeightEleH}) by substituting $r_{d,u}[n]\tan(\theta_{d,u}[n])$ for corresponding $h_{d}[n]$.
Fig. \ref{pathlossFigTheta} shows the curves of $P_{d,u}[n]$ versus $\theta_{d,u}[n]$ under different $r_{d,u}[n]$. 
We can prove that the (\ref{D2UpathlossThetaSubR}) is quasi-convex to $\theta_{d,u}[n]$ and has only one global minimum.
The detail proof can be found in Appendix \ref{appendixA}.
To obtain the optimal $\theta_{d,u}[n]_\mathrm{opt}$ at which $P_{d,u}[n]$ reaches the global minimum, we let the first-order derivation of $P_{d,u}[n]$ to $\theta_{d,u}[n]$ equals zero:
\begin{equation}
\begin{aligned}
\frac{\partial{P_{d,u}[n]}}{\partial{\theta_{d,u}[n]}} & = \frac{20}{\ln(10)}\tan(\theta_{d,u}[n]) \\
& + \frac{ab(\eta_{\mathrm{LoS}} - \eta_{\mathrm{NLoS}})\exp(-b(\theta_{d,u}[n]-a))}{(1 + a\exp(-b(\theta_{d,u}[n] - a)))^2} = 0.
\end{aligned}
\label{FirstDerivationTheta}
\end{equation}
(\ref{FirstDerivationTheta}) is a transcendental equation without closed-form solution. 
However, considering the fact that  (\ref{D2UpathlossThetaSubR}) has only one global minimum which is the single solution for  (\ref{FirstDerivationTheta}), we can further calculate the second-order derivation of $P_{d,u}[n]$ to $\theta_{d,u}[n]$:
\begin{equation}
\begin{aligned}
& \frac{\partial^2{P_{d,u}[n]}}{\partial{\theta_{d,u}[n]}^2} = \frac{20}{\ln(10)}\sec^2(\theta_{d,u}[n]) \\
& \quad \quad \quad \quad + \frac{2a^2b^2(\eta_{\mathrm{LoS}} - \eta_{\mathrm{NLoS}})\exp(-2b(\theta_{d,u}[n]-a))}{(1 + a\exp(-b(\theta_{d,u}[n] - a)))^3} \\
& \quad \quad \quad \quad - \frac{ab^2(\eta_{\mathrm{LoS}} - \eta_{\mathrm{NLoS}})\exp(-b(\theta_{d,u}[n]-a))}{(1 + a\exp(-b(\theta_{d,u}[n] - a)))^2}.
\end{aligned}
\label{SecondDerivationTheta}
\end{equation}
Then, the $\theta_{d,u}[n]_\mathrm{opt}$ can be calculated through the Newton-Raphson method:
\begin{equation}
\begin{aligned}
\theta_{d,u}[n]_{i+1} = \theta_{d,u}[n]_{i} - \frac{{P_{d,u}[n]}^{\prime}(\theta_{d,u}[n]_{i})}{{P_{d,u}[n]}^{\prime\prime}(\theta_{d,u}[n]_{i})}
\end{aligned}
\label{newtonRahoson}
\end{equation}
where the iteration stops when $\theta_{d,u}[n]_{i+1} - \theta_{d,u}[n]_{i} \le \epsilon$ and $\theta_{d,u}[n]_\mathrm{opt} = \theta_{d,u}[n]_{i+1}$.
The calculation of $\theta_{d,u}[n]_\mathrm{opt}$ is constrained by $h_d[n]_\mathrm{d} \le r_{d,u}[n]\tan(\theta_{d,u}[n]) \le h_d[n]_\mathrm{u}$ where $h_d[n]_\mathrm{d}$ and $h_d[n]_\mathrm{u}$ are upper and lower bounds of $h_d[n]$ due to D2B link quality constraint.
After obtaining $\theta_{d,u}[n]_\mathrm{opt}$ for each $\theta_{d,u}[n]$, the optimal $h_d[n]_\mathrm{opt}$ can be calculated as
\begin{equation}
{h_d[n]_\mathrm{opt}} =
  \begin{cases}
    h_d[n]_\mathrm{u}, \quad \theta_{d,u}[n]_\mathrm{opt} \ge \arctan(\frac{h_d[n]_\mathrm{up}}{r_{d,u}[n]}) \\
    h_d[n]_\mathrm{d}, \quad \theta_{d,u}[n]_\mathrm{opt} \le \arctan(\frac{h_d[n]_\mathrm{down}}{r_{d,u}[n]}) \\
    {r_{d,u}[n]}\tan(\theta_{d,u}[n]_\mathrm{opt}), \quad \text{otherwise.}
  \end{cases}
 \label{caseOptHeight}
\end{equation}

\subsection{Protect Distance Constraint}
Given the assumption that all trajectories are closed curves in 3D space, the start slot of each trajectory can be any slots on the trajectory.
Because all trajectories are assumed to have same length of period $N$, for any trajectory, different start slots can lead to different inter-DBS distances at all following slots.
Since that, it is efficient to schedule the start slot of each DBS to prevent violating protect distance constraint (\ref{optimal:l}) in following slots.
We address protect distance constraint in the start slot scheduling process due to two reasons:
First, the feasible set of constraint (\ref{optimal:l}) is non-convex for trajectory related variables, i.e., $\mathbf{L}$ and $\mathbf{H}$, optimizing them in horizontal trajectory or height optimization problems can significantly increase the problem complexity.
Besides, by ensuring the protect distance constraint through start slot scheduling, the DBS trajectory planning result can be maintained, which achieves better average D2U pathloss than modifying those optimized trajectories.
Since the start slot scheduling is not an optimization problem, we can accept any start slots set as long as it ensures constraint (\ref{optimal:l}).
In this work, we apply a greedy-based searching algorithm, as shown in Algorithm \ref{algStartSchedule}, to iteratively schedule the start slots of all DBSs $d_{i}~\forall i = 1, 2, \dots, |\mathcal{D}|$. 
In each iteration, $d_{i}$ sequentially sets its start slot as $n_{j}~\forall j = 1, 2, \dots, N$ and calculate the 3D distances between $d_{i}$ and previous scheduled $d_{k}~\forall k \le i$ at every slots. 
\begin{algorithm}[htbp] 
\caption{Start slots scheduling algorithm} 
\begin{algorithmic}[1]
\State Generate start slots set $\mathcal{S} = \{n_1 = 1, n_2 = 1, \dots, n_{|\mathcal{D}|} = 1\}$ for DBS $d_1, d_2, \dots, d_{|\mathcal{D}|}$.
\For{$i = 1, 2, \dots, |\mathcal{D}|$}
    \For{$j = 1, 2, \dots, N$}
        \State Set $d_{i}$'s start slot as $n_{j}$.
        \State Calculate distance between $d_{i}$ and $d_{k}~\forall k \le i$ at all 
        $~~~~~~~~~$ slots with $d_{i}$ starts at $n_{j}$.
        \State Break if all distances are larger than $Z_\mathrm{min}$
    \EndFor
    \If{$Z_\mathrm{min}$ is violated for all $n_{j} \in N$}
        \State Set $i = 1,~n_1 = n_1 + 1$.
    \EndIf
\EndFor
\end{algorithmic}
\label{algStartSchedule} 
\end{algorithm}
If any start slot $n_{j}$ ensures protect distance constraint at every slots, the start slot of $d_{i}$ is temporally scheduled to $n_{j}$ and break to the $d_{i+1}$ iteration.
If all $n_{j}~\forall j = 1, 2, \dots, N$ on $d_{i}$'s trajectory cannot ensure protect distance constraint, the algorithm abandons the current and all previous scheduled DBSs and re-run the first iteration of $d_{1}$ with updated start slot $n_1 = n_1 + 1$.
Algorithm \ref{algStartSchedule} stops until all $d_{i}$ are scheduled with feasible start slots.

\subsection{Proposed Algorithm}
By decoupling the decision variable set into multiple blocks, i.e., $\mathbf{A}_{t}$, $\mathbf{K}_{t}$, $\mathbf{l}_d[n]~\forall d,n$, $h_d[n]~\forall d,n$, each block's sub-problem can be optimized respectively with other blocks keeping constant.
Therefore, the problem (\ref{optimal}) can be solved through iteratively optimizing those sub-problems until the results converge, which yields to the classic BCD method.
Based on the BCD method, we propose the algorithm to solve the multi-DBS 3D trajectory planning and scheduling problem, which shows in Algorithm \ref{algBCD}.
$\mathbf{A}_{t}$, $\mathbf{K}_{t}$, $\mathbf{W}_{t}$ denote the AoI association, D2U communication scheduling and DBS trajectories after each iteration $t$, respectively. 
$\mathbf{W}_{t}$ is composed by $\mathbf{L}_{t}$ and $\mathbf{H}_{t}$.
According to the BCD method, the proposed algorithm ensures convergence since the global optimal results of all sub-problem are accurately achieved \cite{wu2017joint} \cite{bertsekas1999nonlinear}.

Initial trajectories are required for the first iteration of AoI association.
Without loss of generality, we set the same initial height $h_\mathrm{0} \in [h_d[n]_{d}, h_d[n]_{u}]$ for all DBS. 
For each DBS, we apply a circle initial trajectory with radius $r_\mathrm{0} = 1~\mathrm{m}$.
Given the assumption in subsection A that the initial D2U communication scheduling $k_{d,u}[n]$ equals one for $\forall d,u,n$, it is better to deploy the center of each circle trajectory to the position where the summation of D2U pathloss between its adjacent AoIs (will be associated to the DBS with high probability) is minimized.
Therefore, the classic k-means algorithm can be effectively applied to determine the initial trajectory center of each DBS by substituting D2U pathloss for the geometric distance in original algorithm. 
To reduce the convergence time and improve the result quality, we apply the k-means ++ algorithm which prefers centroid seeds with large mutual distances \cite{arthur2007kmeans++}. 
\begin{algorithm}[htbp] 
\caption{Multi-DBS 3D trajectory planning and scheduling algorithm} 
\begin{algorithmic}[1]
\State Initiate initial U2D communication scheduling $\mathbf{K}_{0}$, initial height $h_\mathrm{0}$.
\State Calculate initial horizontal trajectory $\mathbf{L}_{0}$ through k-means ++ algorithm. 
\State Set $t = 1$, $\Delta{W} = \infty$.
\While{$\Delta{W} \ge \mathbf{\epsilon}$}
  \State Solve problem (\ref{subpAssociation}) to obtain $\mathbf{A}_{t}$ by treating $\mathbf{K}_{t-1}$, $\mathbf{L}_{t-1}$ 
  $~~~~$ and $\mathbf{H}_{t-1}$ as constants.
  \State Solve problem (\ref{subpScheduling}) to obtain $\mathbf{K}_{t}$ by treating $\mathbf{A}_{t}$, $\mathbf{L}_{t-1}$ 
  $~~~~$ and $\mathbf{H}_{t-1}$ as constants.
  \For{$d \in \mathcal{D}, n = 1,2,\dots,N$}
        \State Solve problem (\ref{subpHorTrajEleR}) to obtain $\mathbf{L}_{d}[n]_\mathrm{opt}$ by treating 
        $~~~~~~~~~$ $\mathbf{A}_{t}$, $\mathbf{K}_{t}$, $\mathbf{H}_{t-1}$ and $\mathbf{L}_{d}[\bar{n}]~\forall d,\bar{n} \neq n$ as constants.
        \State Update $\mathbf{L}_{t}$ with $\mathbf{l}_{d}[n]_\mathrm{opt}$.
  \EndFor
  \For{$d \in \mathcal{D}, n = 1,2,\dots,N$}
        \State Solve problem (\ref{subpHeightEleH}) to obtain $h_{d}[n]_\mathrm{opt}$ by treating 
        $~~~~~~~~~$ $\mathbf{A}_{t}$, $\mathbf{K}_{t}$, $\mathbf{L}_{t}$ and $h_{d}[\bar{n}]~\forall d, \bar{n} \neq n$ as constants. 
        \State Update $\mathbf{H}_{t}$ with $h_{d}[n]_\mathrm{opt}$.
  \EndFor
  \State Update $\mathbf{W}_{t}$ with $\mathbf{L}_{t}$ and $\mathbf{H}_{t}$.
  \State $t = t + 1$.
  \State $\Delta{W} = \mathbf{W}_{t} - \mathbf{W}_{t-1}$.
\EndWhile
\State Run Algorithm \ref{algStartSchedule} to ensure protect distance constraint.
\end{algorithmic}
\label{algBCD} 
\end{algorithm}

\section{Numerical Results}
We conduct extensive simulations to verify the performance of our proposed algorithm in minimizing average D2U pathloss of the network. 
The simulations are link level without simulating specific MAC or upper layers protocols.
The BS is located at the origin point (coordinate $(0,0,0)$) and the side-length of grid is set to $20~\mathrm{m}$.
Both U2D and D2B pathloss models are configured in suburban scenario.
To provide additional spectrum resources for DA-RAN and reduce the interference to terrestrial RAN users, the frequency band of D2U communication $f_{c}$ is expected to be different from the licensed cellular band.
Like most commercial drone products \cite{dji2018mavic} \cite{cheng2018air} and DBS related works \cite{yaliniz2019spatial} \cite{Dhekne2016} \cite{mozaffari2017mobile} \cite{zhao2018deployment}, we use the $2.4~\mathrm{GHz}$ unlicensed band as the carrier for D2U communications.
D2B communications use the $850~\mathrm{MHz}$ LTE band according to the D2B pathloss model \cite{al2017modeling}.
By allocating different carrier frequencies, the interference between D2U and D2B communication can be prevented.
Initial height $h_\mathrm{0}$ is set to $80~\mathrm{m}$ within the working-zone of DBS over the whole BS radio coverage area \cite{shi2018multiple}. 
We treat $\delta_{t}$ as the minimal time unit to calculate related variables including $V_\mathrm{max}$, $H_\mathrm{max}$, etc. 
There is no need to assign specific value for $\delta_{t}$ in the simulation, however, according to the general specifications of commercial drones ($50-70~\mathrm{km}/\mathrm{h}$ for horizontal speed, $3-5~\mathrm{m}/\mathrm{s}$ for ascent/descent speed) \cite{tseng2017autonomous}, the approximate value of $\delta_{t}$ is around $10~\mathrm{s}$.
Table. \ref{Table_simulation} shows detail simulation parameters. 
\begin{table}[htpb]
\centering
\caption{Simulation Parameters}
\label{Table_simulation}
\begin{tabular}{ll}
\hline\noalign{\vskip 0.3mm}\hline\noalign{\smallskip}
Parameters & Numerical Values\\
\noalign{\smallskip}\hline\noalign{\smallskip}
BS radio coverage radius $r_{\mathrm{BS}}$ & $900~\mathrm{m}$\\
AoI number $|\mathcal{U}|$ & $20$\\
D2U parameters $(\eta_{\mathrm{LoS}},\eta_{\mathrm{NLoS}},a,b)$ & (0.1,21,4.88,0.43)\\
D2B parameters $(\alpha,A,\theta_{0},B,\eta_{0})$ & (3.04,-23.29,-3.61,4.14,20.7)\\
Carrier frequencies (D2U, D2B) & $(2.4~\mathrm{GHz}, 850~\mathrm{MHz})$\\
Slots amount in one period $N$ & $60~\mathrm{slot}$\\
D2B pathloss constraint $P_{\mathrm{DB}}$ & $80~\mathrm{dB}$\\
Minimal per-AoI slot number $S_\mathrm{min}$ & $10~\mathrm{slot}$\\
Maximal per-DBS AoI number ${|\mathcal{A}_d|}_\mathrm{max}$ & $6$\\
Maximal horizontal speed $V_\mathrm{max}$ & $30,50,70,90,110~\mathrm{m}/\mathrm{slot}$\\
Maximal vertical speed $H_\mathrm{max}$ & $10~\mathrm{m}/\mathrm{slot}$\\
Protect Distance $Z_\mathrm{min}$ & $200~\mathrm{m}$\\
Trajectory difference $\mathbf{\epsilon}$ & $0.1~\mathrm{m}$ for each slot \\
\hline\noalign{\vskip 0.3mm}\hline
\end{tabular}
\vspace{-0.3cm}
\end{table}

\begin{figure}[htbp]
  \centering
  \subfloat[3D view]{\includegraphics[width=0.37\textwidth]{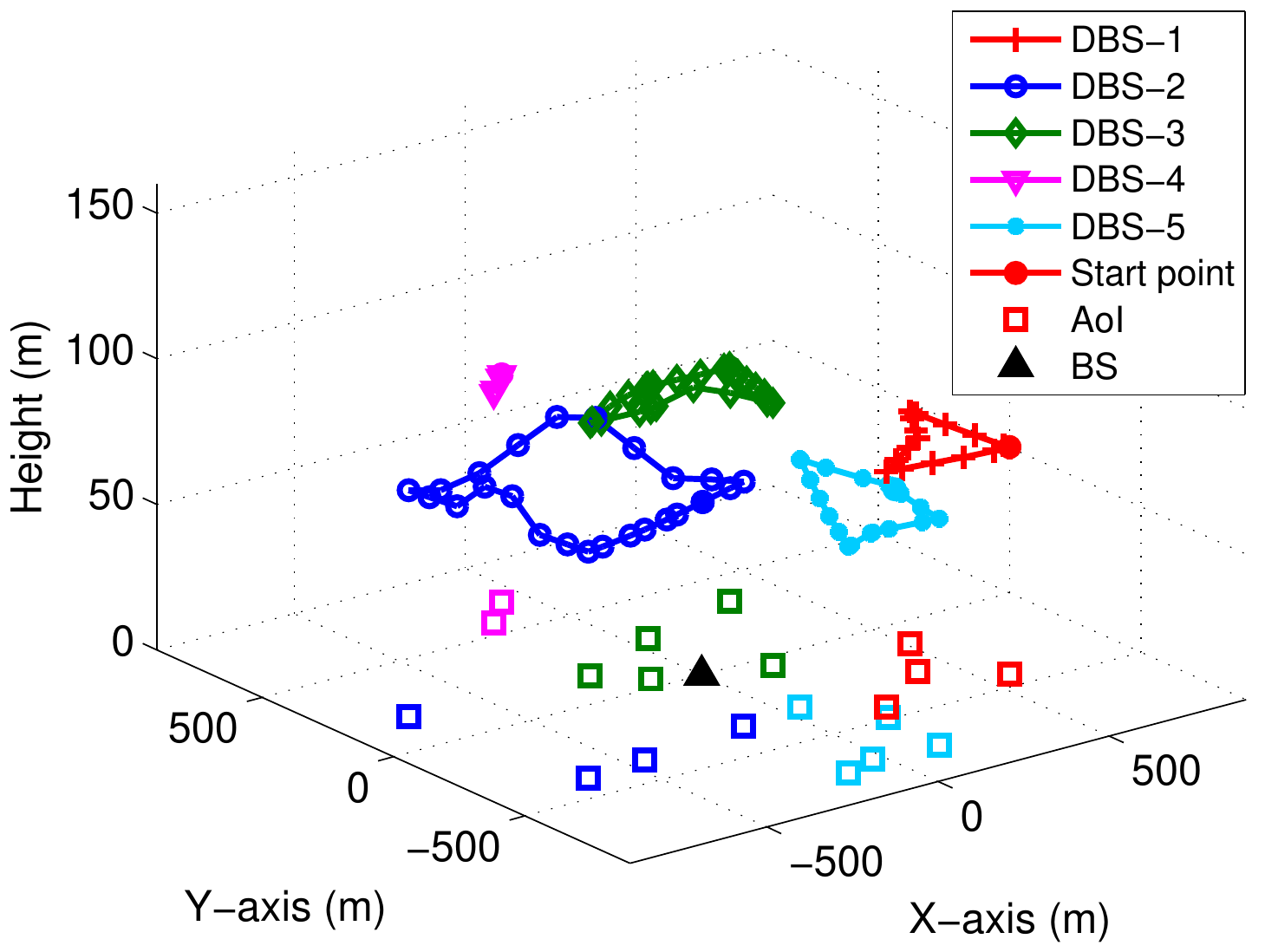}}\\
  \subfloat[2D projection on X-Y plane]{\includegraphics[width=0.37\textwidth]{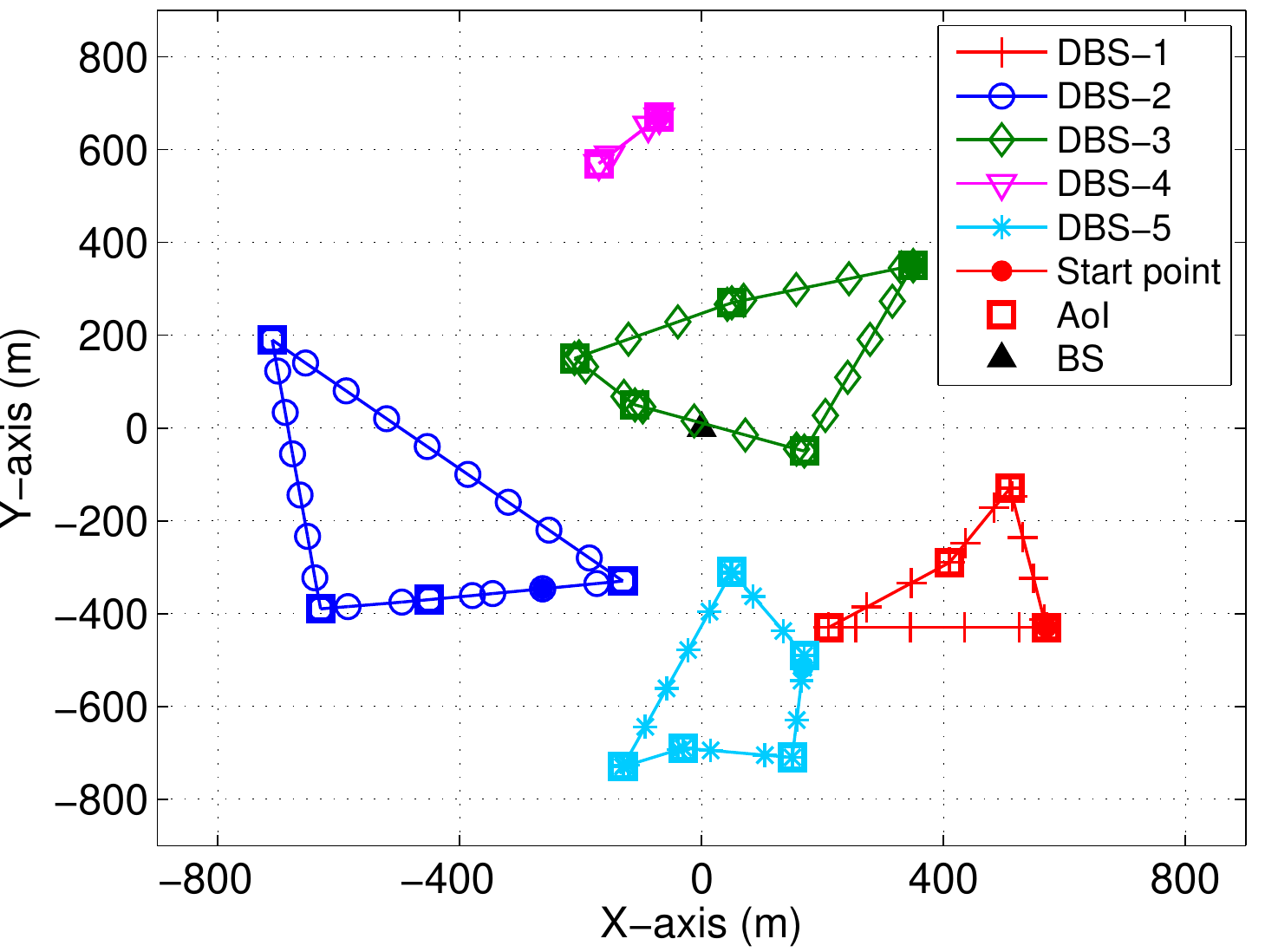}}\\
  \subfloat[Heights]{\includegraphics[width=0.37\textwidth]{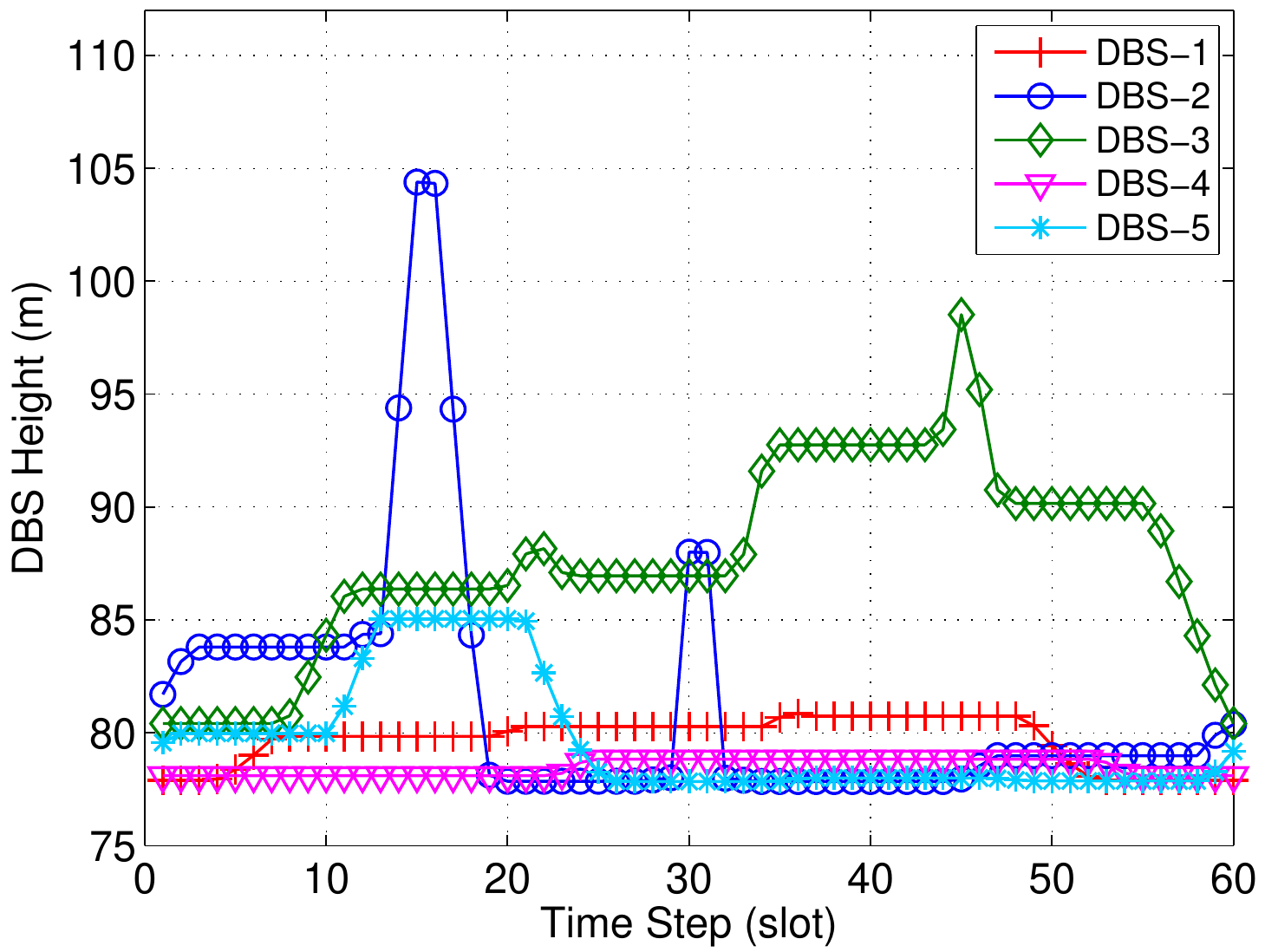}}\\
  \subfloat[Protect distance guarantee]{\includegraphics[width=0.37\textwidth]{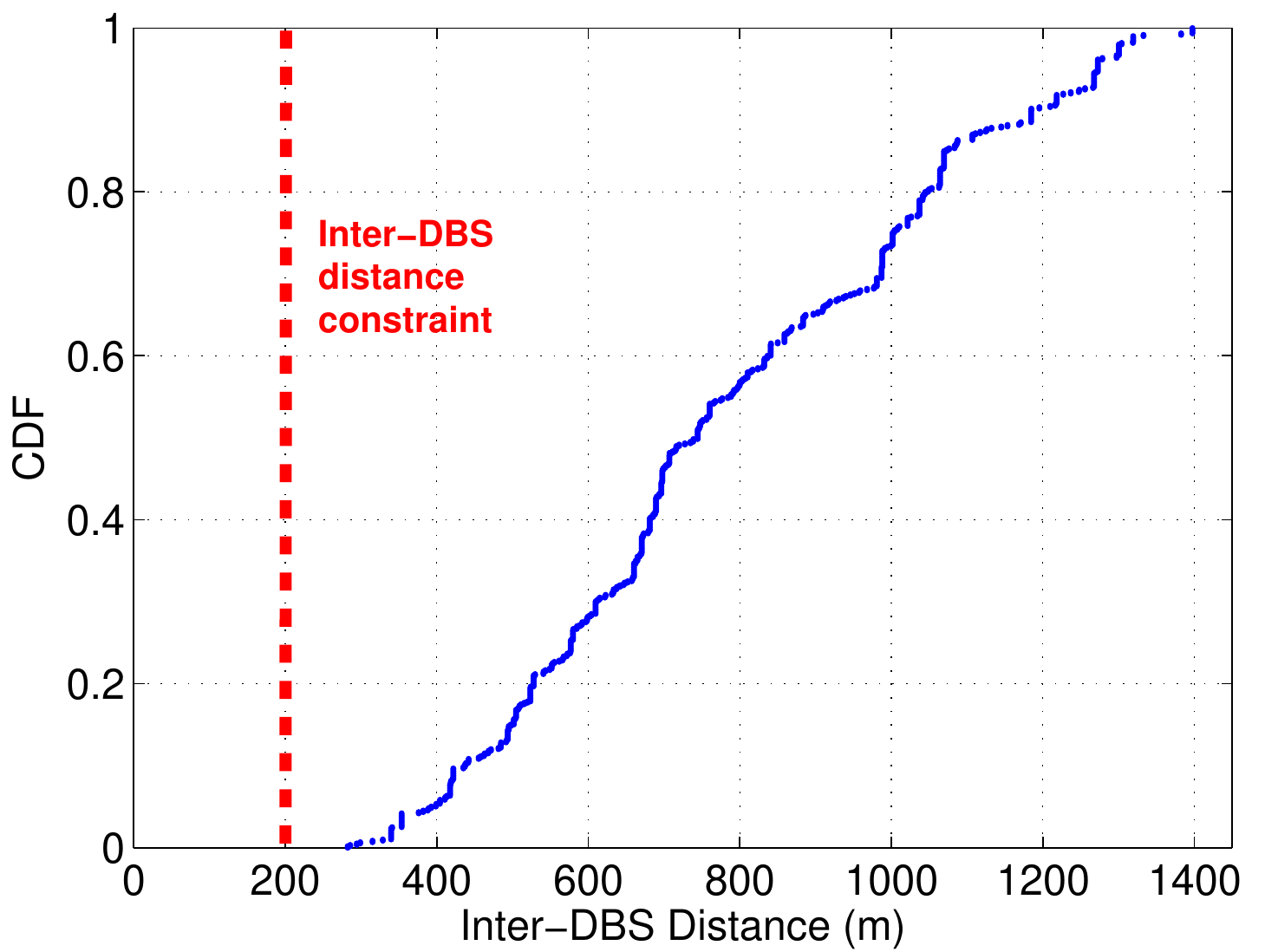}}
  \caption{Trajectory planning results of 5 DBSs serving 20 AoIs.}
  \label{disFigall}
  \vspace{-0.5cm}
\end{figure}

\subsection{3D Trajectory Planning for Multiple DBSs}
Fig. \ref{disFigall} shows the scenario where the trajectories of five DBSs are optimized to serve twenty AoIs with $V_\mathrm{max} = 90~\mathrm{m}/\mathrm{slot}$. 
The closed curves dotted by different markers in Fig. \ref{disFigall}(a) and Fig. \ref{disFigall}(b) denote different DBS trajectories; the squares on the X-Y plane represent AoIs.
AoIs are associated to corresponding DBSs with same colors. 
Fig. \ref{disFigall}(c) illustrates the changes of flying height within one period.
As shown in Fig. \ref{disFigall}(a) and \ref{disFigall}(b), for each DBS, the optimized trajectory can fly over all its associated AoIs and form a closed curve in 3D space.
In Fig. \ref{disFigall}(c), all DBS flying height curves are lower bounded around $78~\mathrm{m}$, which is the lower bound of $h_d[n]_{d}$ due to the D2B pathloss constraint.
\begin{figure*}[htbp]
  \centering
  \subfloat[$V_\mathrm{max} = 30~\mathrm{m}/\mathrm{slot}$]{\includegraphics[width=0.245\textwidth,height=0.16\textheight]{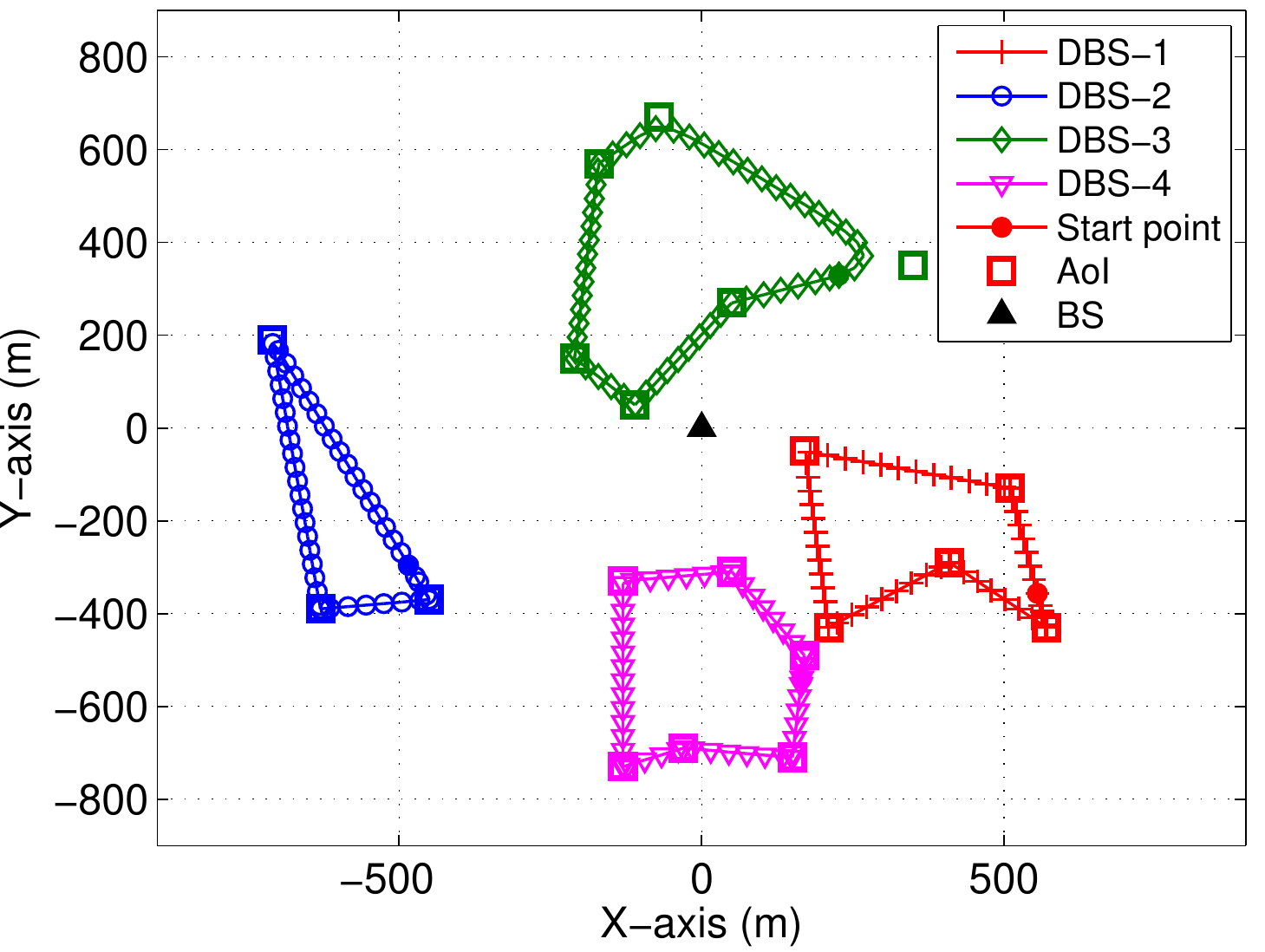}}
  \subfloat[Heights of $V_\mathrm{max} = 30~\mathrm{m}/\mathrm{slot}$]{\includegraphics[width=0.245\textwidth,height=0.16\textheight]{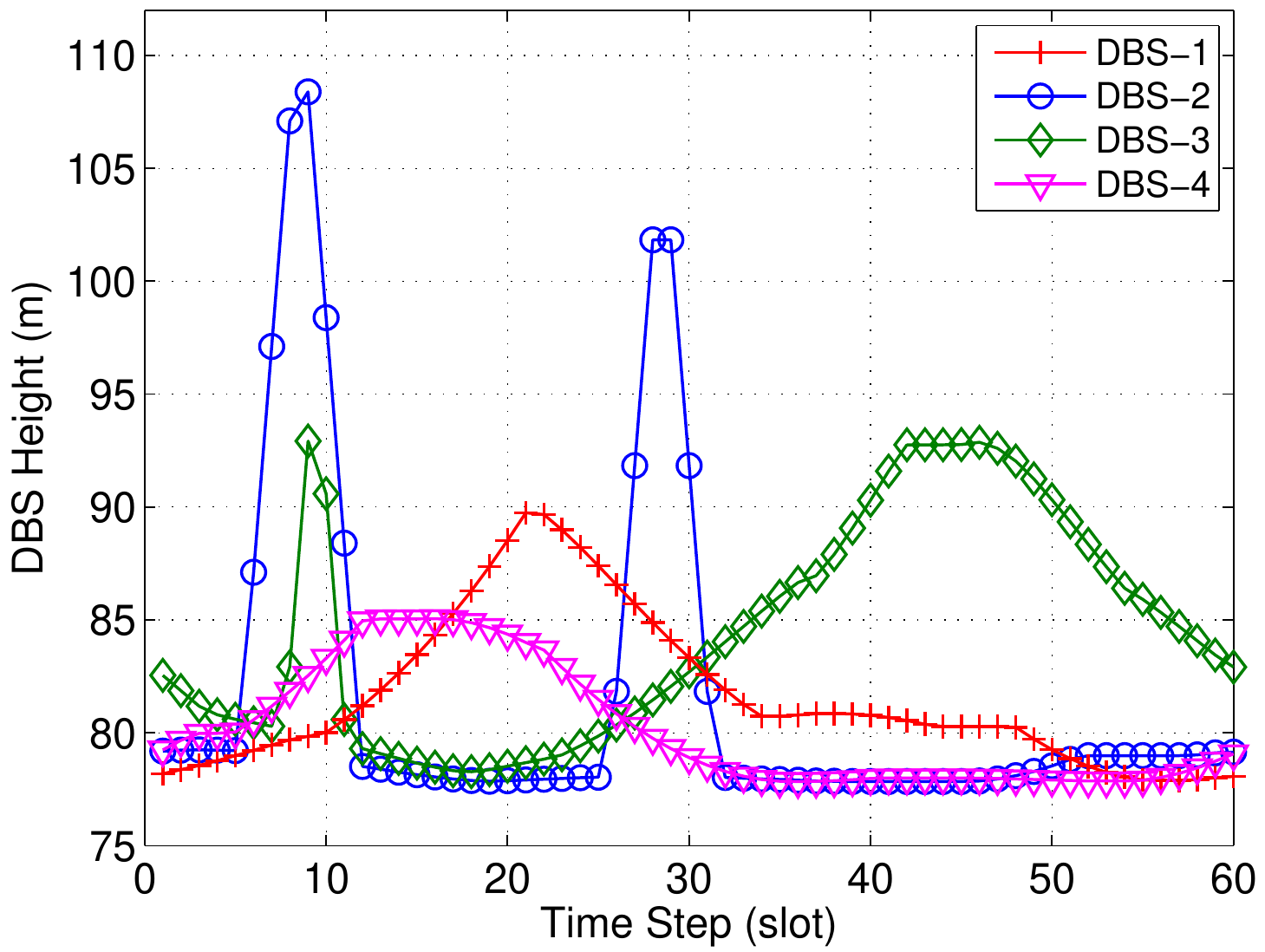}}
  \subfloat[$V_\mathrm{max} = 110~\mathrm{m}/\mathrm{slot}$]{\includegraphics[width=0.245\textwidth,height=0.16\textheight]{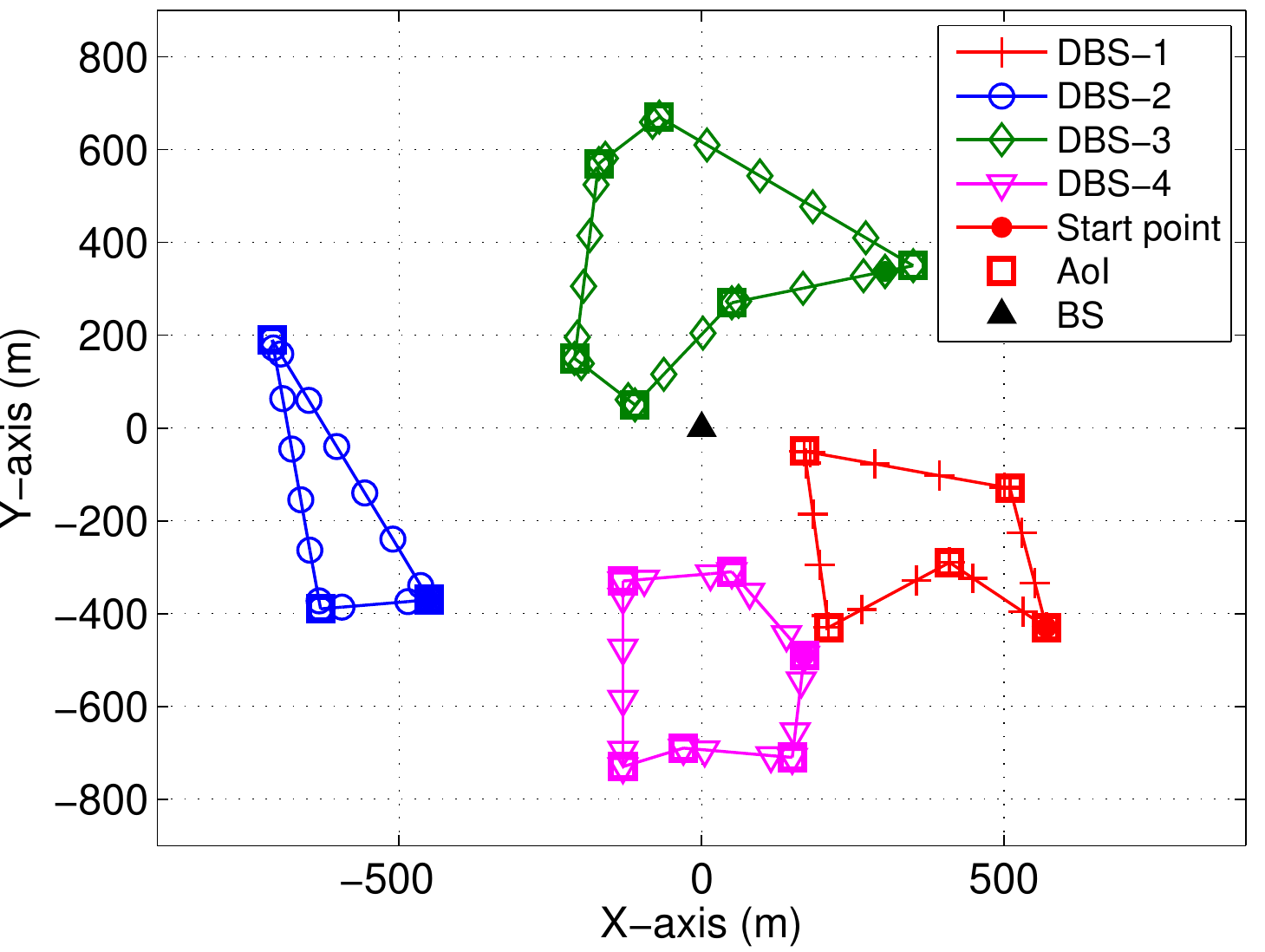}}
  \subfloat[Heights of $V_\mathrm{max} = 110~\mathrm{m}/\mathrm{slot}$]{\includegraphics[width=0.245\textwidth,height=0.16\textheight]{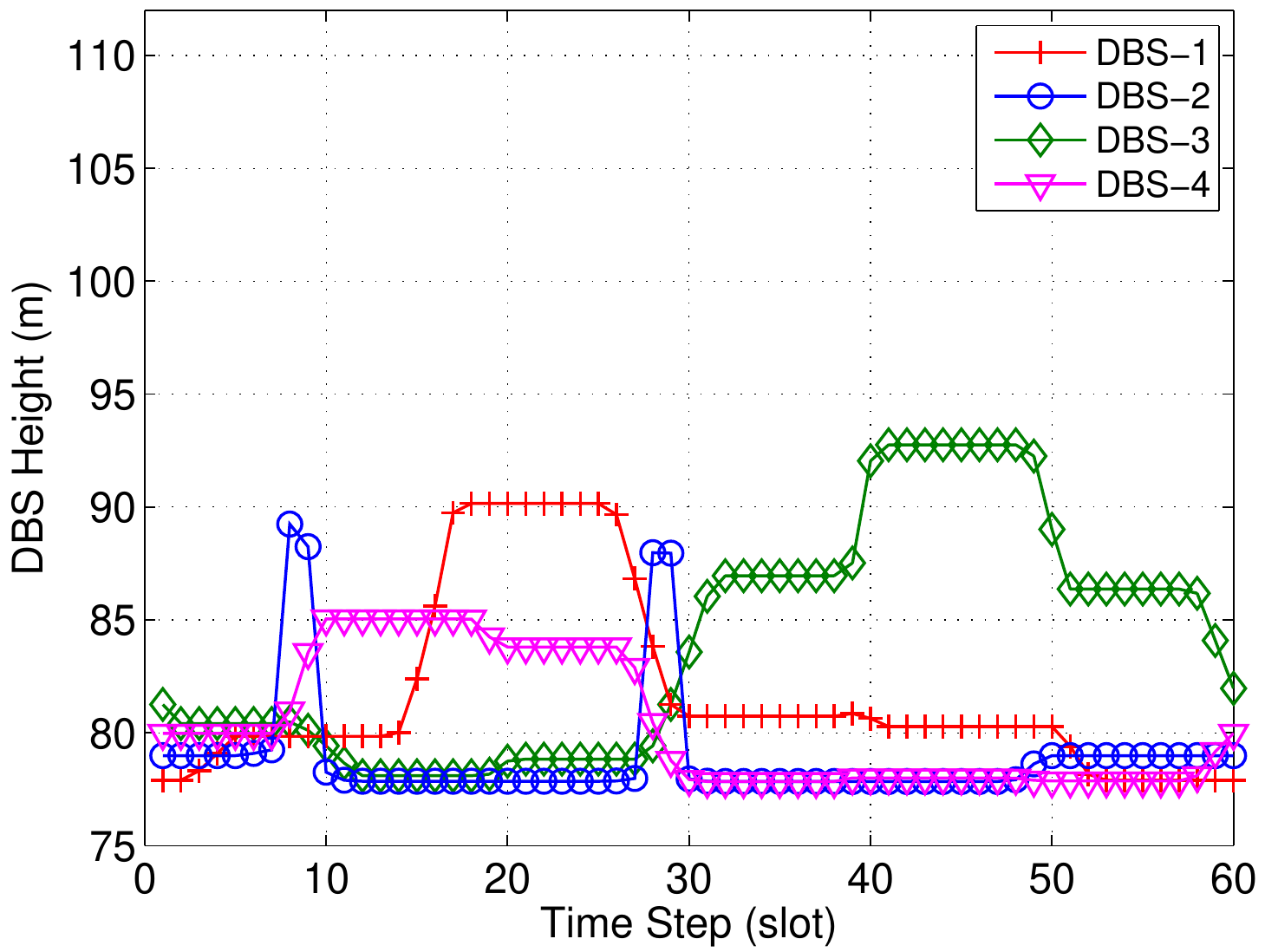}}
  \caption{Trajectory planning results impacted by horizontal speeds.}
  \label{speedImpact}
  \vspace{-0.6cm}
\end{figure*}
\begin{figure*}[htbp]
  \centering
  \subfloat[$|\mathcal{D}| = 4$]{\includegraphics[width=0.245\textwidth,height=0.16\textheight]{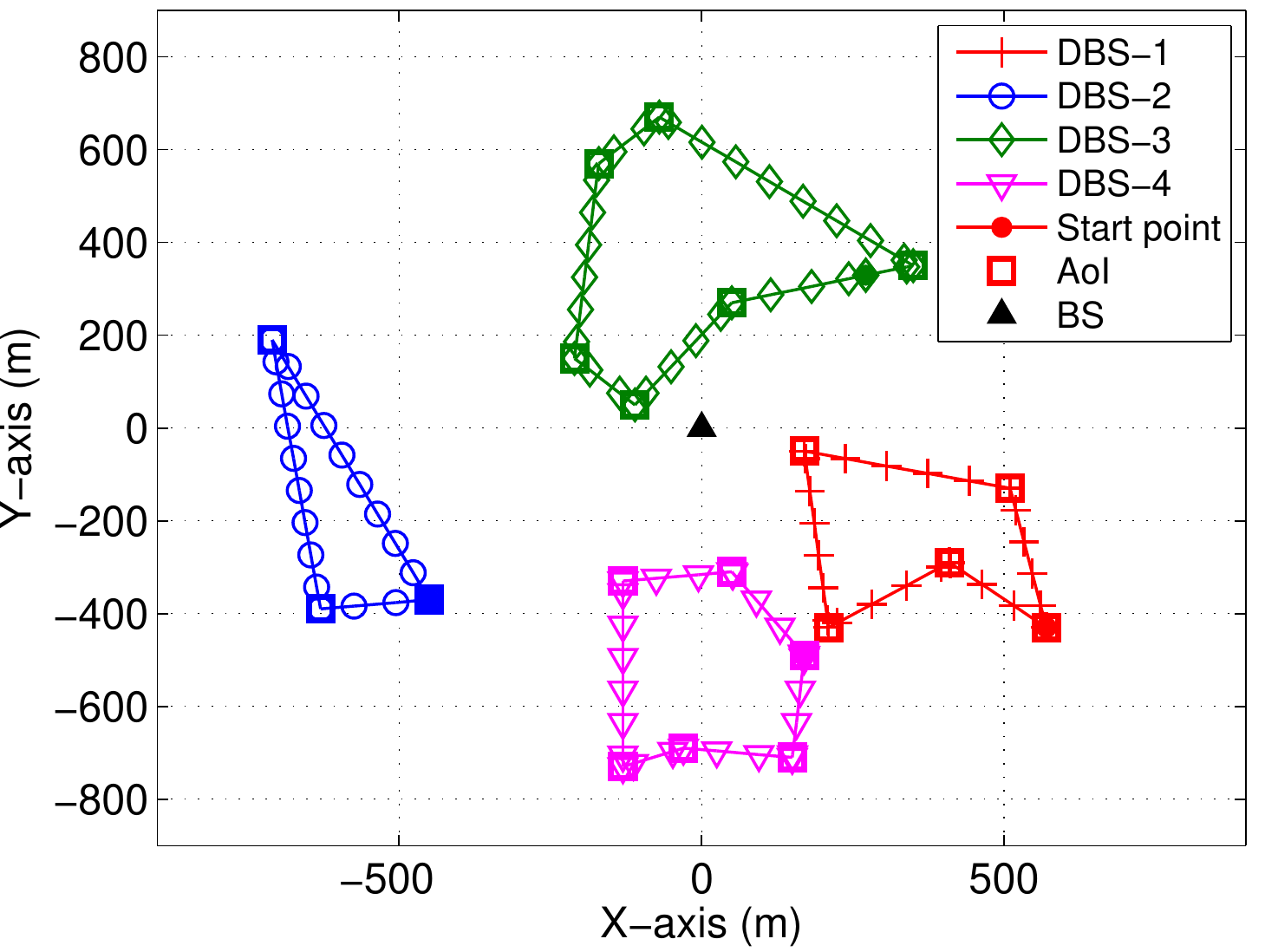}}
  \subfloat[Heights of $|\mathcal{D}| = 4$]{\includegraphics[width=0.245\textwidth,height=0.16\textheight]{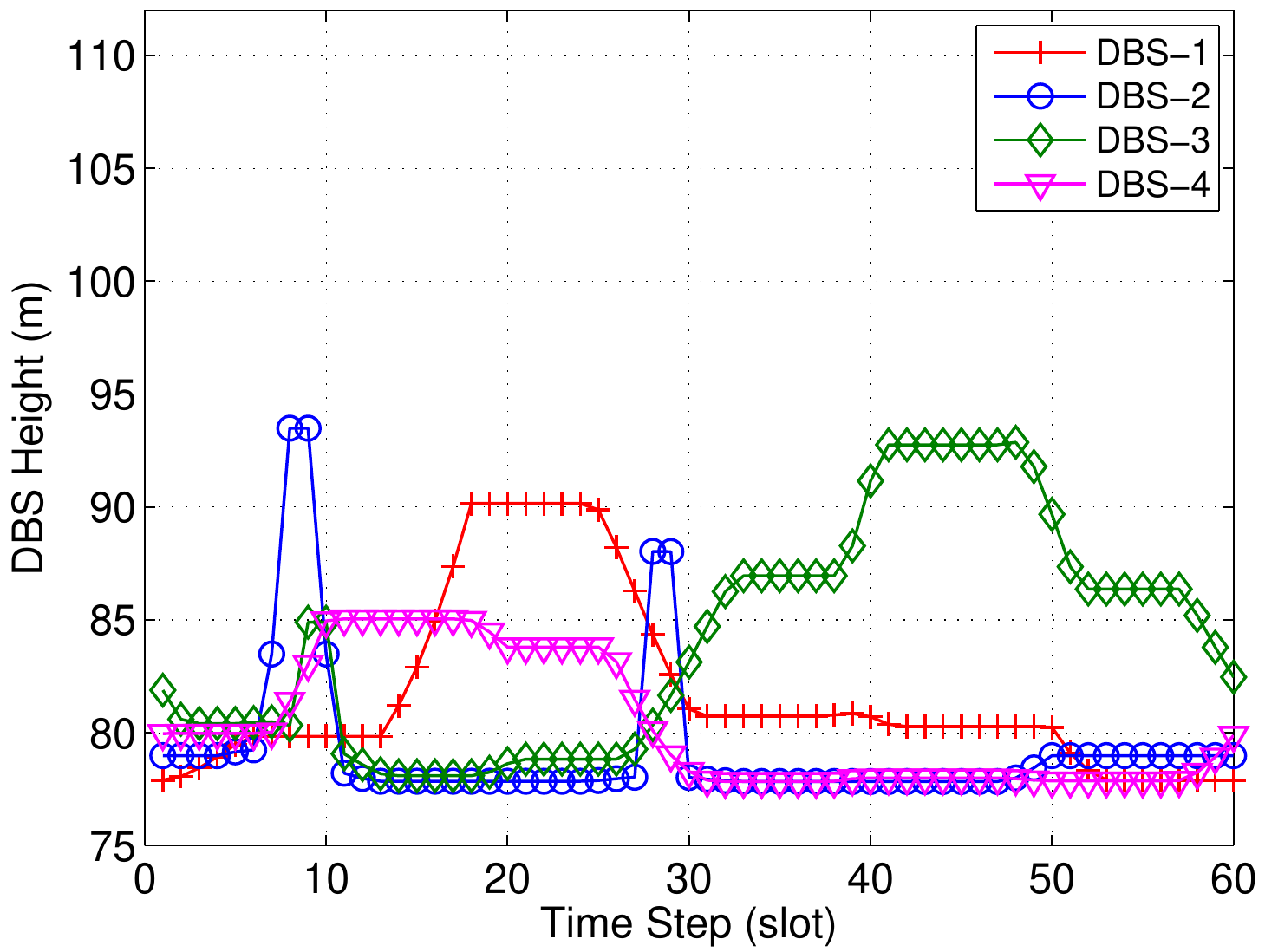}}
  \subfloat[$|\mathcal{D}| = 7$]{\includegraphics[width=0.245\textwidth,height=0.16\textheight]{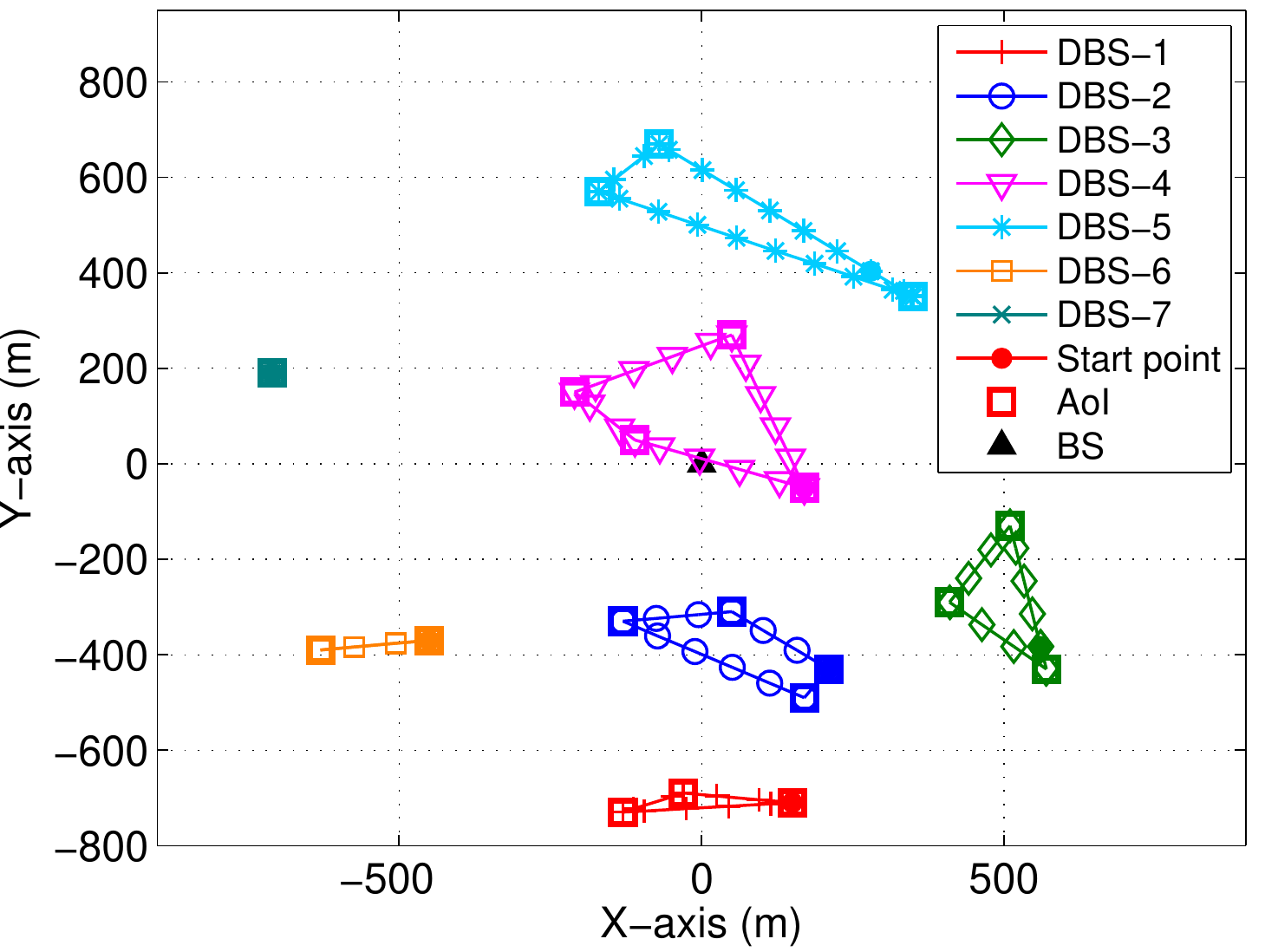}}
  \subfloat[Heights of $|\mathcal{D}| = 7$]{\includegraphics[width=0.245\textwidth,height=0.16\textheight]{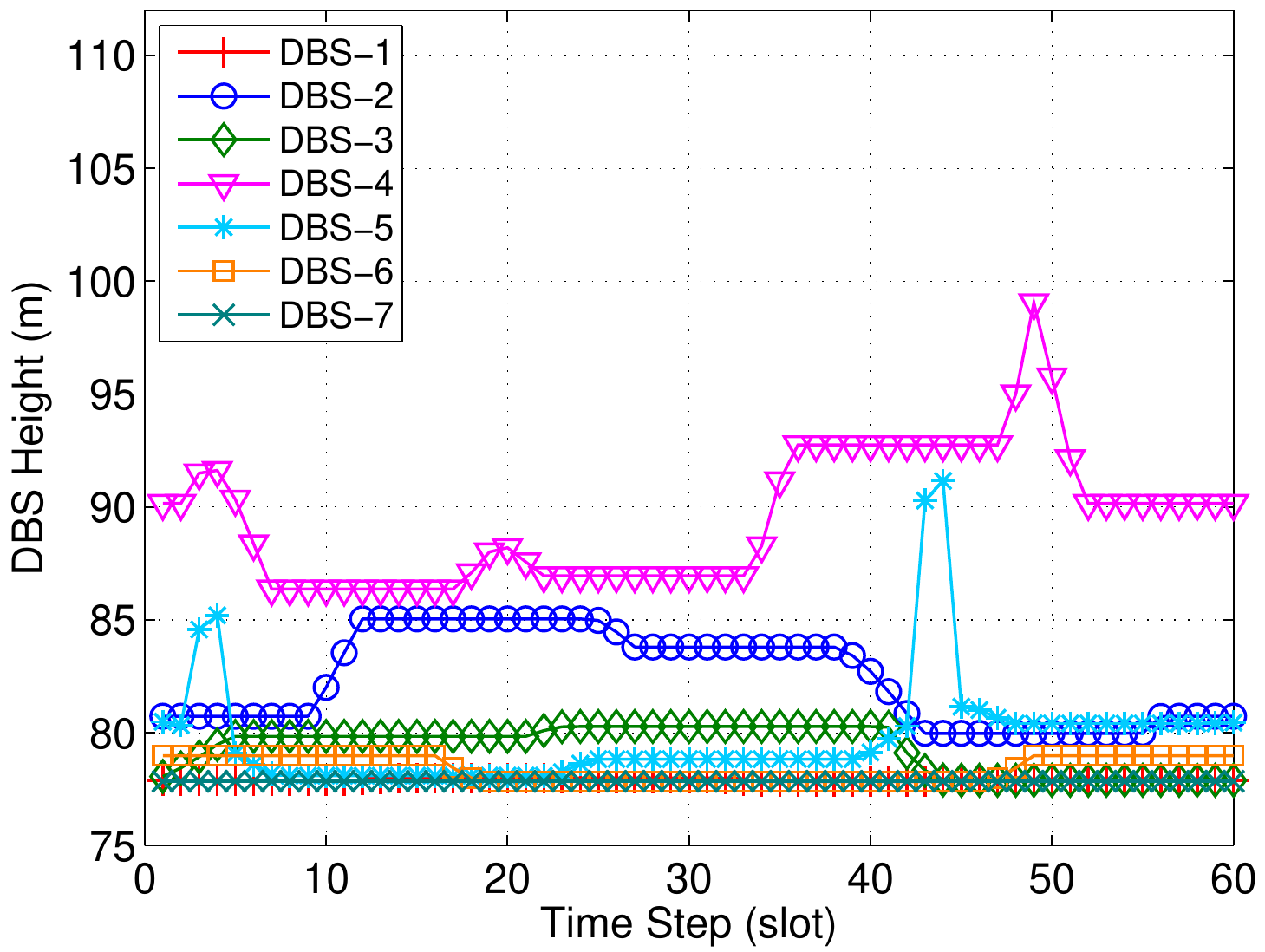}}
  \caption{Trajectory planning results impacted by DBS number.}
  \label{DBSnumImpact}
  \vspace{-0.5cm}
\end{figure*}

Note that for each trajectory in Fig. \ref{disFigall}(b), the summation of dots on the section between two AoIs is less than $50\%$ of the total slots number $N$.
Besides, the height curves in Fig. \ref{disFigall}(c) show the trend to maintain fixed values for consecutive slots.
Combining Figs. \ref{disFigall}(b) and \ref{disFigall}(c), we can justify that the remaining dots in Fig. \ref{disFigall}(b) are overlapped above the associated AoIs, and those overlapped dots corresponds to the consecutive slots have fixed heights in Fig. \ref{disFigall}(c).
In other word, the proposed algorithm prefers hovering DBSs above the associated AoIs, while leave minimal slots for the travelling process between adjacent hovering positions. 
Such a ``hovering effect'' can be explained as follows.
For any slot $n$, the proposed algorithm is prone to small $r_{d,u}[n]$ which minimizes the D2U pathloss with given $h_d[n]$.
On the contrary, the smaller the $r_{d,u}[n]$ is, the higher the probability that $r_{d,u}[n]\theta_{d,u}[n]_\mathrm{opt} \le h_d[n]_\mathrm{d}$.
Based on  (\ref{caseOptHeight}), if $r_{d,u}[n]\theta_{d,u}[n]_\mathrm{opt} \le h_d[n]_\mathrm{d}$, the optimal height equals the lower bound of flying height $h_d[n]_\mathrm{d}$ at current position.
Since the minimal $r_{d,u}[n]$ equals zero, the minimal average D2U pathloss can be achieved by the trajectory with most slots hovering above the AoIs.
From Fig. \ref{disFigall}(c), we can see that several height bursts occur when each DBS is flying between two AoI with a long inter-AoI distance.
At those slots, the DBS is relatively far from the scheduled AoI and  ${r_{d,u}[n]}\tan(\theta_{d,u}[n]_\mathrm{opt})$ can fall in the feasible height range between $h_d[n]_{d}$ and $h_d[n]_{u}$ constrained by D2B pathloss threshold.
Therefore, the optimal heights prefer to approach the value of ${r_{d,u}[n]}\tan(\theta_{d,u}[n]_\mathrm{opt})$ in those slots, which leads to the height bursts.
Fig. \ref{disFigall}(d) shows the cumulative distribution function (CDF) of inter-DBS distance within one period $T$.
The red dotted line is the protect distance constraint.
It can be seen that all inter-DBS distances of the final trajectory planning result are larger than the protect distance threshold $Z_\mathrm{min}$, which indicates the effectiveness of the start slots scheduling algorithm to ensure protect distance constraint.

\begin{figure}[htbp]
  \centering
  \includegraphics[width=0.37\textwidth]{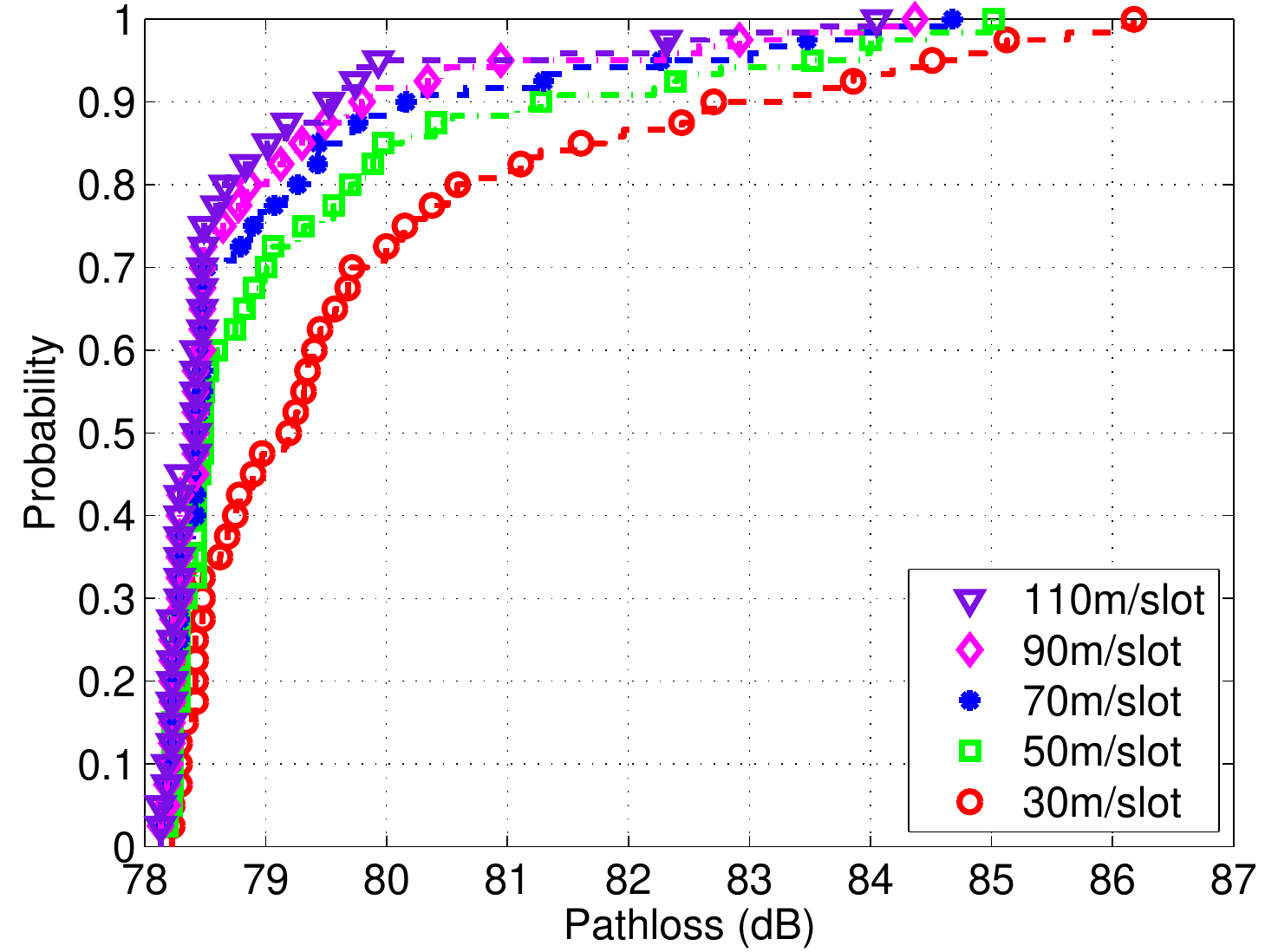}
  \caption{CDF of D2U pathloss under different $V_\mathrm{max}$.}
  \label{pdfSpeed}
  \vspace{-0.4cm}
\end{figure}

Fig. \ref{speedImpact} presents two groups of trajectory planning results with $V_\mathrm{max} = 30~\mathrm{m}/\mathrm{slot}$ and $V_\mathrm{max} = 110~\mathrm{m}/\mathrm{slot}$, respectively. 
The available DBS number $|\mathcal{D}|$ equals four for both groups.
As shown in Fig. \ref{speedImpact}, the AoI associations are same under different $V_\mathrm{max}$.
The trajectories in Fig. \ref{speedImpact}(a) cannot fly over every associated AoI since the maximal horizontal speed is too small to ensure the DBSs to approach every associated AoIs within one period.
In Fig. \ref{speedImpact}(c) where $V_\mathrm{max}$ is high enough, the DBSs can even hovering on each associated AoI for few slots since the flying interval between two hovering positions requires less slots.
Comparing Fig. \ref{speedImpact}(b) and \ref{speedImpact}(d), we can see that the variation of flying height with $V_\mathrm{max} = 30~\mathrm{m}/\mathrm{slot}$ is larger than the height variation with  $V_\mathrm{max} = 110~\mathrm{m}/\mathrm{slot}$.
Because in low $V_\mathrm{max}$ scenario, the travelling process between two AoIs requires more slots than that in the high $V_\mathrm{max}$ scenario, the optimal height ${r_{d,u}[n]}\tan(\theta_{d,u}[n]_\mathrm{opt})$ in small $V_\mathrm{max}$ has higher probability to fall into the feasible flying height range. 
Considering the four DBSs trajectory planning scenario, the CDF of D2U pathloss under different $V_\mathrm{max}$ are compared in Fig. \ref{pdfSpeed}. 
Given any pathloss threshold, We can see that the probability of D2U pathloss less than the threshold raises as the $V_\mathrm{max}$ increases.

\begin{figure}[htbp]
  \centering
  \includegraphics[width=0.37\textwidth]{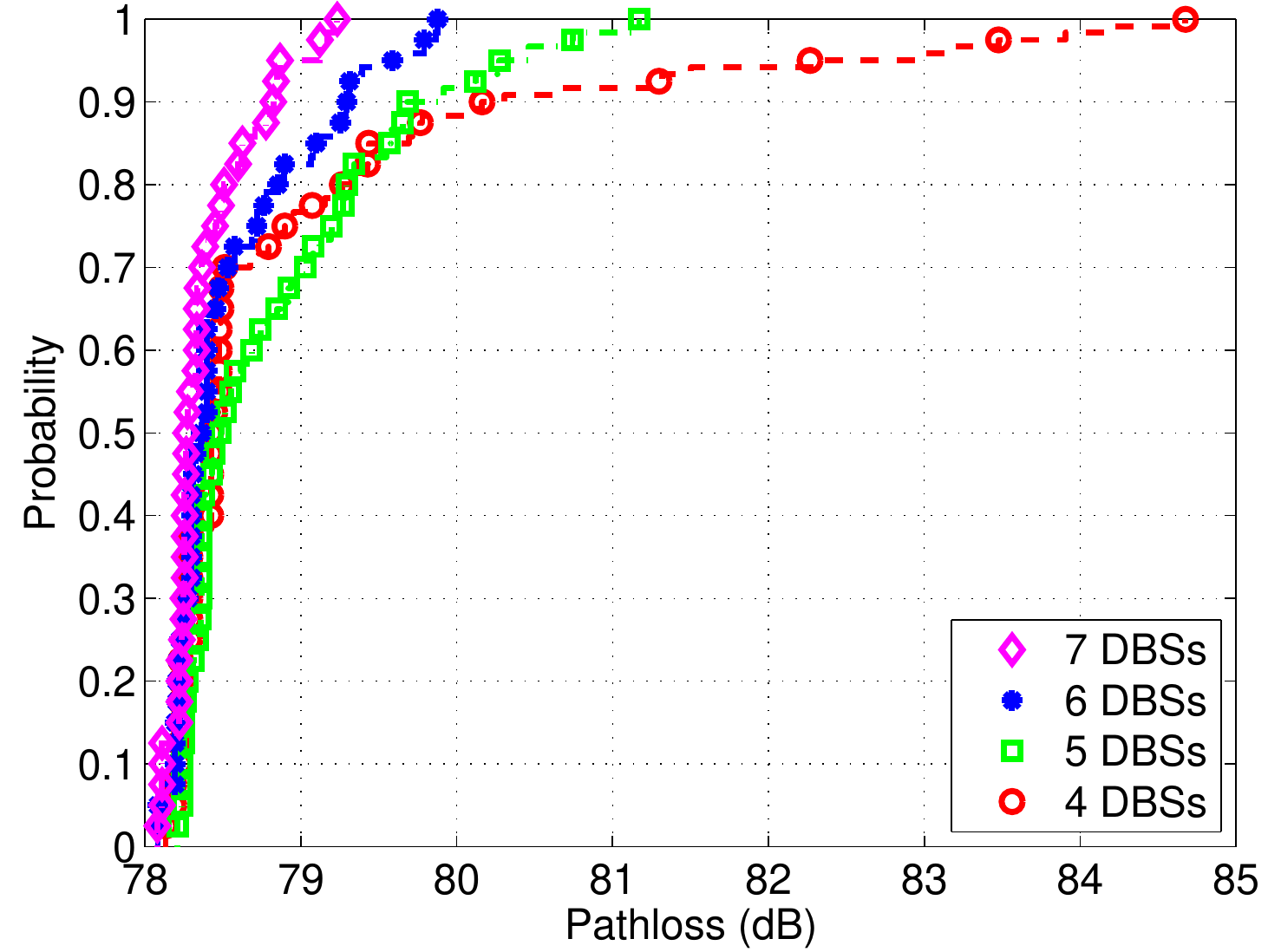}
  \caption{CDF of D2U pathloss under different $|\mathcal{U}|$.}
  \label{pdfNumber}
  \vspace{-0.5cm}
\end{figure}

Fig. \ref{DBSnumImpact} compares the trajectory planning results when $|\mathcal{D}| = 4$ and $|\mathcal{D}| = 7$.
The horizontal speed is set as $V_\mathrm{max} = 70~\mathrm{m}/\mathrm{slot}$ for both scenarios.
As the number of available DBS increases, the average number of AoI associated to one DBS is reduced, some trajectories can even degenerate to one static deployment position when the corresponding DBS is associated with only one AoI.
On the other hand, since the average $r_{d,u}[n]$ length is also reduced with the decreasing of associated AoI number for each DBS, the variation of flying height can be reduced with the increasing of $|\mathcal{D}|$.
Fig. \ref{pdfNumber} shows the CDF of D2U pathloss under different $|\mathcal{D}|$.
Similar to Fig. \ref{pdfSpeed}, the probability of D2U pathloss less than any given threshold increases as more numbers of DBS are provided.

\begin{figure}[htbp]
  \centering
  \includegraphics[width=0.37\textwidth]{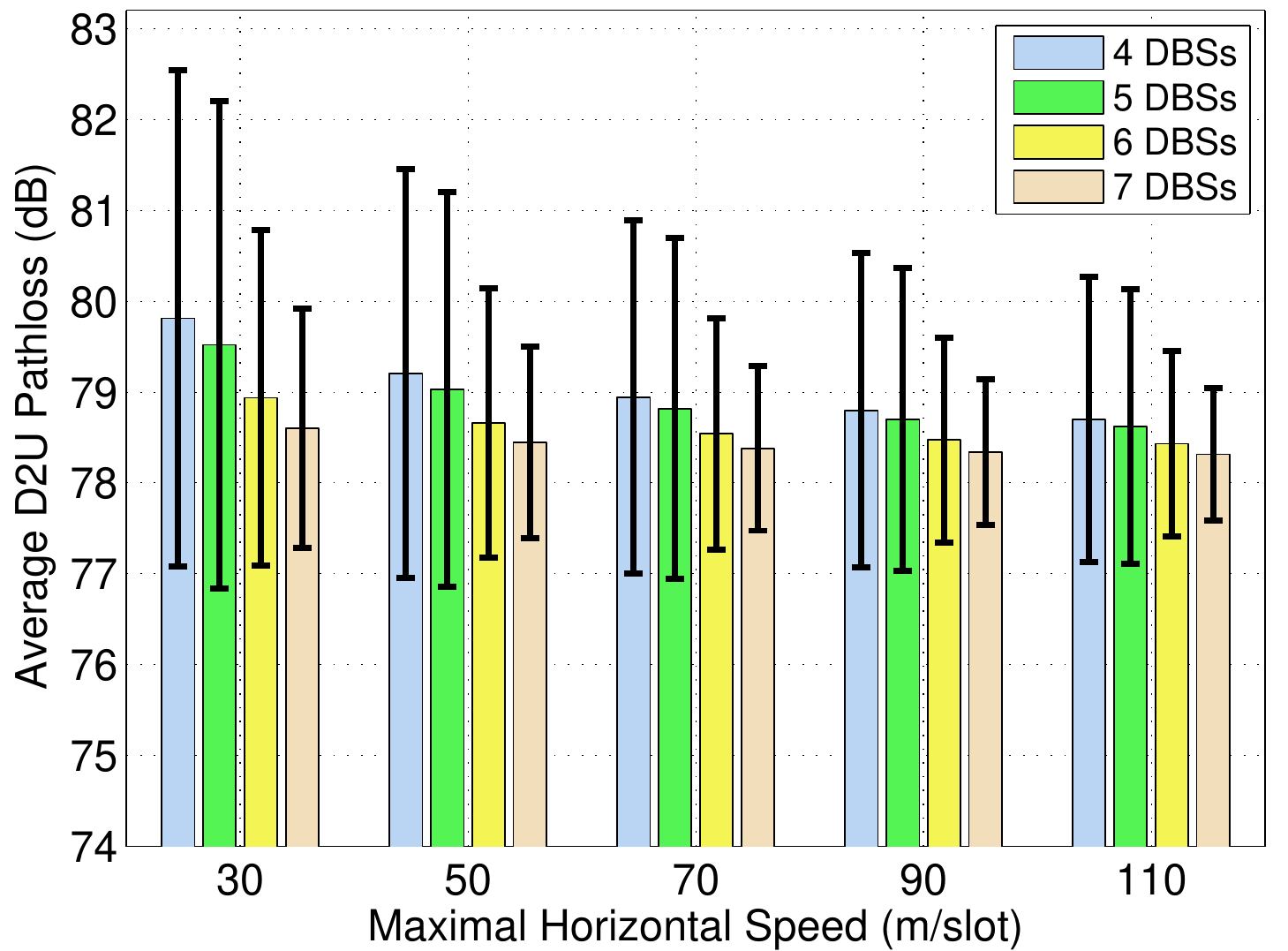}
  \caption{Average D2U pathloss with different horizontal speeds and DBS numbers.}
  \label{trajPerformCmp}
  \vspace{-0.5cm}
\end{figure}

Fig. \ref{pdfSpeed} and Fig. \ref{pdfNumber} indicate that the D2U pathloss performance can be influenced by both $V_\mathrm{max}$ and $|\mathcal{D}|$.
In Fig. \ref{trajPerformCmp}, we further investigate the average pathloss performance with different $V_\mathrm{max}$ and $|\mathcal{D}|$.
From preceding analyses, we know that both the higher $V_\mathrm{max}$ and larger $|\mathcal{D}|$ can lead to smaller average $r_{d,u}$, which eventually reduces the average pathloss level.
According to Fig. \ref{trajPerformCmp}, given the same number of DBS, the average pathloss level decreases slightly as the $V_\mathrm{max}$ increases; while significant average pathloss level reduction occurs as the $|\mathcal{D}|$ increases under fixed maximal horizontal speed.
Therefore, we can conclude that both raising the horizontal speed and increasing the number of available DBS can promote the average D2U pathloss performance of DBS trajectory planning, while increasing the number of available DBSs is proved to be more efficient than raising horizontal flying speed.
Note that this conclusion is valid for the average D2U pathloss performance of the whole network only. 
Since the standard deviations of pathloss (error-bars) plotted in Fig. \ref{trajPerformCmp} are highly overlapped with each other, the D2U pathloss performance of specific DBS-to-AoI pair can vary a lot. 
Nevertheless, as the horizontal speed and available DBS number increase, the standard deviation is reduced, which indicates that the user fairness can also be promoted by raising $V_\mathrm{max}$ and $|\mathcal{D}|$.

\begin{figure}[htbp]
  \centering
  \includegraphics[width=0.37\textwidth]{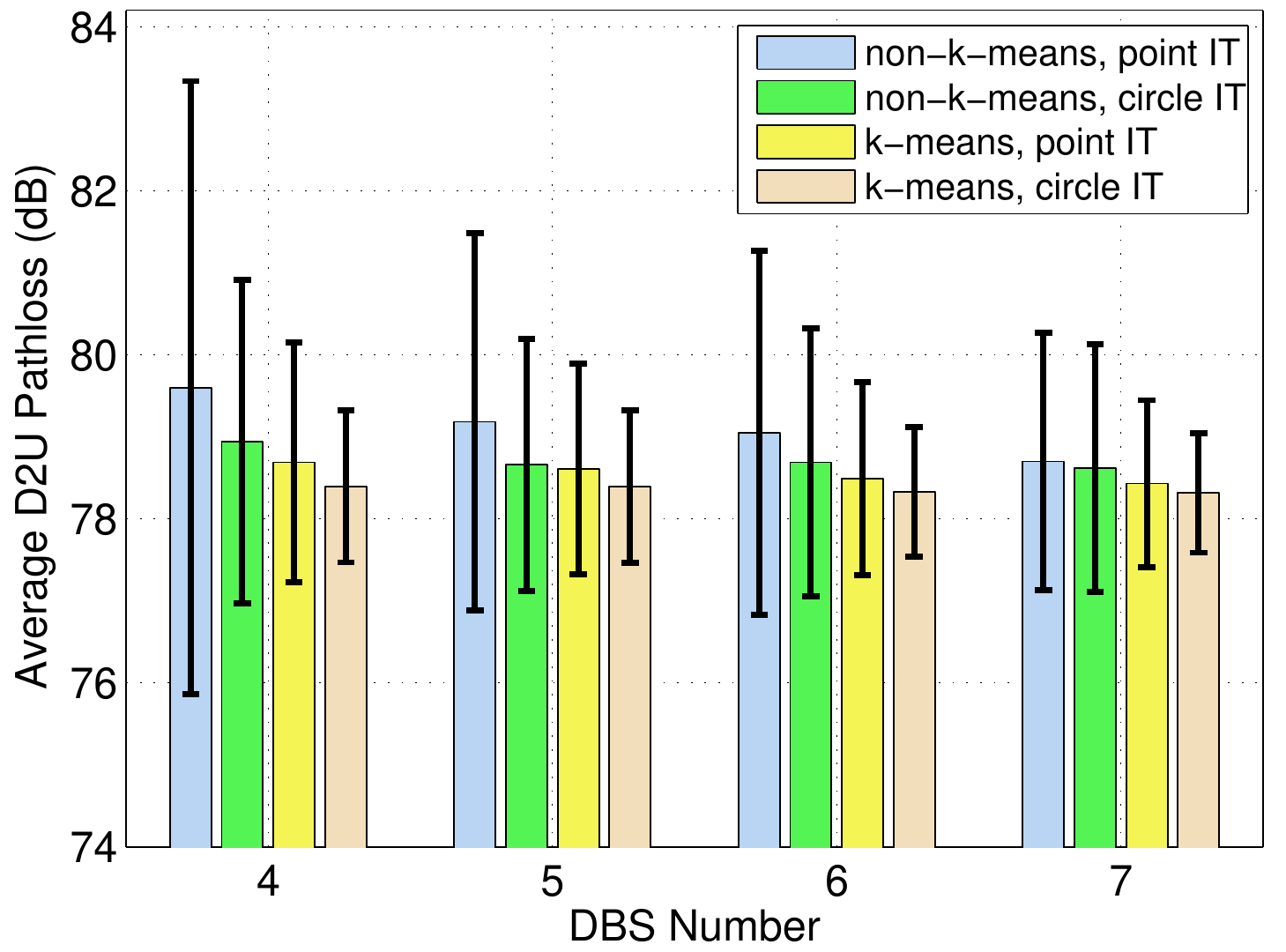}
  \caption{Initial trajectory comparison.}
  \label{initialTrajImpacts}
  \vspace{-0.5cm}
\end{figure}

We further analyze the impacts of different initial trajectories (ITs) to the achieved average D2U pathloss performance.
We compare four types of ITs, i.e., 1) circle IT with the center location determined by k-means ++; 2) point IT (where the trajectory shrinks to one hovering point) with the point location determined by k-means ++; 3) circle IT with the center location uniformly distributed over ${|\mathcal{S}|}_{\mathrm{bs}}$; 4) point IT with the point location uniformly distributed over ${|\mathcal{S}|}_{\mathrm{bs}}$.
Fig. \ref{initialTrajImpacts} shows the comparison result.
We can see that the k-means-based circle IT used in the proposed algorithm achieves minimal average D2U pathloss and pathloss standard deviation. 
Comparing the performance gaps between ITs, we note that applying k-means ++ algorithm has more significant impact to both average D2U pathloss and pathloss standard deviation than using circle-shaped IT.

\subsection{Performance Comparison}
\begin{figure}[htbp]
  \centering
  \includegraphics[width=0.37\textwidth]{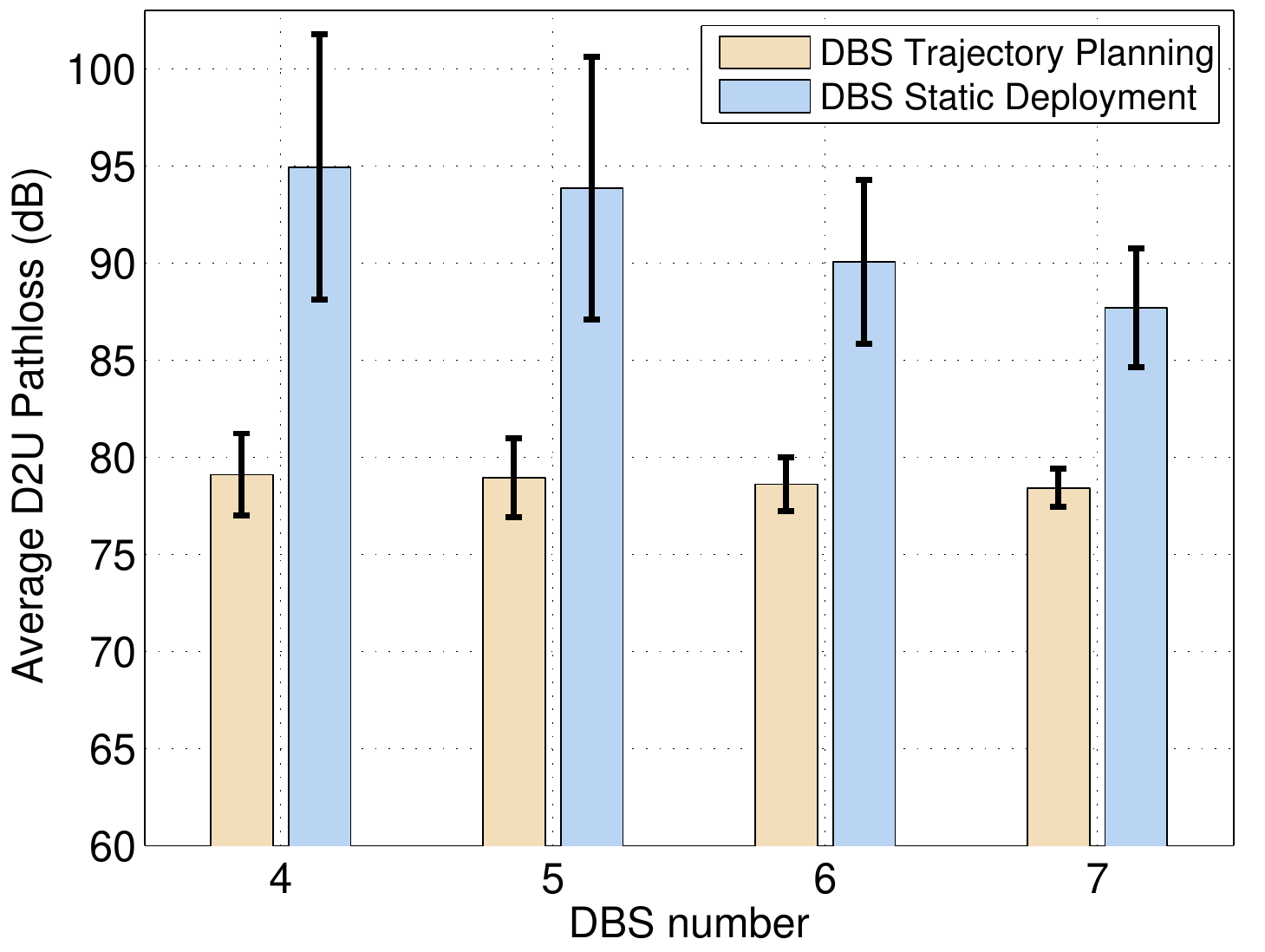}
  \caption{Average D2U pathloss comparison between trajectory planning and static deployment of multiple DBSs.}
  \label{trajPsoCmp}
  \vspace{-0.5cm}
\end{figure}

To highlights the efficiency of the proposed multi-DBS 3D trajectory planning and scheduling, we compare the average D2U pathloss performance achieved by our proposed trajectory planning algorithm, as well as the static DBS deployment scheme in Fig. \ref{trajPsoCmp} and Table \ref{Table_deviation}. 
For static DBS deployment algorithm, we use the per-drone iterated particle swarm optimization (DI-PSO) algorithm proposed in \cite{shi2018multiple}.
Without loss of generality, we use all data achieved by $V_\mathrm{max} = 30,50,70,90,110~\mathrm{m}/\mathrm{slot}$ to calculate the average D2U pathloss of the proposed trajectory planning algorithm.
From Fig. \ref{trajPsoCmp}, we can see that the average D2U pathloss of both algorithms are reduced as the available DC number increases. 
However, the D2U pathloss performance achieved by our trajectory planning algorithm maintains $10-15~\mathrm{dB}$ smaller than that achieved by the DI-PSO algorithm.

The user fairness promotion provided by the proposed trajectory planning algorithm is indicated by the error-bars in Fig. \ref{trajPsoCmp} and the D2U pathloss standard deviation comparison in Table \ref{Table_deviation}.
$\sigma_t$ and $\sigma_s$ are D2U pathloss standard deviations for DBS trajectory planning and static DBS deployment, respectively.
In Fig. \ref{trajPsoCmp}, we can see that the standard deviation of D2U pathloss achieved by both algorithms are reduced as the available DBS number increases. 
While the $\sigma_t$ maintains less than half of the $\sigma_s$, there is no overlap between the error-bars achieved by two algorithms.
From Table \ref{Table_deviation}, we can calculate that the DBS trajectory planning can lower the D2U pathloss standard deviation by $68.34\%$ on average compared with the static DBS deployment.

To highlight the cost-efficiency of the trajectory planning algorithm, and provide a guideline to determine the number of required DBS, we compare the minimal required DBS number under different D2U pathloss thresholds in Table \ref{Table_minNumDBS}. 
The D2U pathloss threshold is a strict constraint, which means that for any DBS at any slot, the D2U pathloss to any AoI cannot exceed it.
As shown in Table \ref{Table_minNumDBS}, under the same D2U pathloss threshold, the minimal required DBS amount is decreased as the maximal horizontal speed of DBS increases.
Besides, given any threshold levels, the static DBS deployment always requires two to four more DBSs than the DBS trajectory planning algorithm, which implies that the DBS trajectory planning algorithm is more economical than the static DBS deployment.
\begin{table}[htpb]
\centering
\caption{D2U pathloss standard deviation comparison}
\label{Table_deviation}
\begin{tabular}{l|l|l|l|l}
\hline\noalign{\vskip 0.3mm}\hline
DBS number & $4$ & $5$ & $6$ & $7$ \\
\hline
$\sigma_t$ & $2.1203$ & $2.0476$ & $1.3923$ & $0.9887$ \\
\hline
$\sigma_s$ & $6.8313$ & $6.7562$ & $4.2329$ & $3.0530$ \\
\hline
$({\sigma_s} - {\sigma_t})/{\sigma_s}$ & $68.96\%$ & $69.69\%$ & $67.11\%$ & $67.61\%$ \\
\hline\noalign{\vskip 0.3mm}\hline
\end{tabular}
\vspace{-0.3cm}
\end{table}
\begin{table}[htpb]
\centering
\caption{Minimal DBS amounts comparison}
\label{Table_minNumDBS}
\begin{tabular}{l|l|l|l|l|l}
\hline\noalign{\vskip 0.3mm}\hline
D2U pathloss threshold ($\mathrm{dB}$) & $98$ & $95$ & $92$ & $89$ & $86$\\
\hline
$30\mathrm{m/slot}$ & $3$ & $4$ & $5$ & $6$ & $8$\\
\hline
$30\mathrm{m/slot}$ & $3$ & $4$ & $5$ & $6$ & $8$\\
\hline
$70\mathrm{m/slot}$ & $3$ & $4$ & $4$ & $6$ & $7$\\
\hline
$90\mathrm{m/slot}$ & $3$ & $4$ & $4$ & $5$ & $7$\\
\hline
$110\mathrm{m/slot}$ & $3$ & $3$ & $4$ & $4$ & $6$\\
\hline\noalign{\vskip 0.3mm}\hline
Static deployment & $6$ & $7$ & $8$ & $10$ & $10$\\
\hline\noalign{\vskip 0.3mm}\hline
\end{tabular}
\vspace{-0.3cm}
\end{table}

\section{Conclusion}
In this paper, we have studied the 3D trajectory planning and scheduling for multiple DBSs in DA-RAN with the state-of-the-art D2U and D2B pathloss models considered.
We have formulated the MINLP problem to minimize the average D2U pathloss achieved by multi-DBS trajectory planning and scheduling. 
To solve the MINLP problem, we have decoupled the MINLP problem into multiple solvable sub-problems, and devised a BCD-based multi-DBS 3D trajectory planning and scheduling algorithm in which the AoI association, U2D communication scheduling, horizontal trajectories, and flying altitudes of DBSs are iteratively optimized.
A start slot scheduling algorithm and a k-means-based circle IT have been proposed to ensure the protect distance constraint and generate initial DBS trajectories.
We have investigated the impacts of available DBS number, horizontal speed and different IT on the achieved average D2U pathloss.
Simulation results have shown that the proposed algorithm can achieve $10-15~\mathrm{dB}$ average D2U pathloss reduction, and promote pathloss standard deviation by $68\%$ when compared with the static DBS deployment algorithm.
In future works, we will investigate the communication and resource allocation of multiple DBS in DA-RAN based on the optimized trajectories.


\ifCLASSOPTIONcaptionsoff
\newpage
\fi
\begin{appendices}
\section{}
Since the first two parts of (\ref{D2UpathlossThetaSubR}) are constant or the function of $r_{d,u}[n]$, analyzing the convexity of (\ref{D2UpathlossThetaSubR}) is equal to analyzing the convexity of (\ref{D2UpathlossThetaSubOnlyTheta}).
(\ref{D2UpathlossThetaSubOnlyTheta}) can be regarded as the summation of two functions:
\begin{equation}
\begin{aligned}
& F(\theta_{d,u}[n]) = F_{1}(\theta_{d,u}[n]) + F_{2}(\theta_{d,u}[n]) \\
& \quad = 20\log(\sec(\theta_{d,u}[n]) + F_{2}(\theta_{d,u}[n]) \\
& \quad = 20\log(\sec(\theta_{d,u}[n]) + \frac{\eta_{\mathrm{LoS}} - \eta_{\mathrm{NLoS}}}{1 + a\exp(-b(\theta_{d,u}[n] - a))}.
\end{aligned}
\label{D2UpathlossThetaSubOnlyTheta:apd}
\end{equation}
The first-order derivations of $F_{1}(\theta_{d,u}[n])$ and $F_{2}(\theta_{d,u}[n])$ are
\begin{subequations}
\begin{align}
& F_{1}^{\prime}(\theta_{d,u}[n]) = \frac{20}{\ln(10)}\tan(\theta_{d,u}[n]), \label{FirstOrderApd:a}\\
& F_{2}^{\prime}(\theta_{d,u}[n]) = \frac{ab(\eta_{\mathrm{LoS}} - \eta_{\mathrm{NLoS}})\exp(-b(\theta_{d,u}[n]-a))}{(1 + a\exp(-b(\theta_{d,u}[n] - a)))^2}. \label{FirstOrderApd:b}
\end{align}
\label{FirstOrderApd}
\end{subequations}
We further calculate the second-order derivations of $F_{1}(\theta_{d,u}[n])$ and $F_{2}(\theta_{d,u}[n])$ as
\begin{subequations}
\begin{align}
& F_{1}^{\prime\prime}(\theta_{d,u}[n]) = \frac{20}{\ln(10)}\sec^2(\theta_{d,u}[n]), \label{SecondOrderApd:a}\\
& F_{2}^{\prime\prime}(\theta_{d,u}[n]) = \frac{ab^2(\eta_{\mathrm{NLoS}} - \eta_{\mathrm{LoS}})\exp(-b(\theta_{d,u}[n]-a))}{(1 + a\exp(-b(\theta_{d,u}[n] - a)))^2} \nonumber\\
& \quad \quad \quad - \frac{2a^2b^2(\eta_{\mathrm{NLoS}} - \eta_{\mathrm{LoS}})\exp(-2b(\theta_{d,u}[n]-a))}{(1 + a\exp(-b(\theta_{d,u}[n] - a)))^3}. \label{SecondOrderApd:b}
\end{align}
\label{SecondOrderApd}
\end{subequations}
Note that (\ref{SecondOrderApd:a}) is always larger than zero for all $\theta_{d,u}[n] \in [0^{\circ},90^{\circ}]$, so (\ref{FirstOrderApd:a}) is proved to be non-decreasing function.
Let (\ref{SecondOrderApd:b}) equals zero, we can calculate the only root of (\ref{SecondOrderApd:b}) $\theta_{d,u}[n]_\mathrm{root} = a + \ln(a)/b$ at which (\ref{FirstOrderApd:b}) achieves its global minimum. 

Given the suburban scenario parameters $(\eta_{\mathrm{LoS}},\eta_{\mathrm{NLoS}},a,b) = (0.1,21,4.88,0.43)$, we can calculate that:
\begin{subequations}
\begin{align}
& F_{1}^{\prime}(90^{\circ}) + F_{2}^{\prime}(90^{\circ}) \ge 0, \label{SecondOrderSumApd:a}\\
& F_{1}^{\prime}(\theta_{d,u}[n]_\mathrm{root}) + F_{2}^{\prime}(\theta_{d,u}[n]_\mathrm{root}) \le 0, \label{SecondOrderSumApd:b}\\
& F_{1}^{\prime}(0^{\circ}) + F_{2}^{\prime}(0^{\circ}) \le 0. \label{SecondOrderSumApd:c}
\end{align}
\label{SecondOrderSumApd}
\end{subequations}
which indicates $F_{1}^{\prime}(\theta_{d,u}[n]) + F_{2}^{\prime}(\theta_{d,u}[n])$, i.e., (\ref{FirstDerivationTheta}), has only one root between $\theta_{d,u}[n]_\mathrm{root}$ and $90^{\circ}$.
Therefore, we can conclude that (\ref{D2UpathlossThetaSubOnlyTheta}) is a uni-modal function with only one global minimum for all $\theta_{d,u}[n] \in [0^{\circ},90^{\circ}]$.

Define the $\theta_{d,u}[n]$ achieves the global minimum of (\ref{D2UpathlossThetaSubOnlyTheta}) as $\theta_{d,u}[n]_\mathrm{opt}$, $a \in [0^{\circ},90^{\circ}]$, $b \in [0^{\circ},90^{\circ}]$ and $t = \lambda{a} + (1-\lambda)b~\forall \lambda \in [0,1]$.
If $a \le b \le \theta_{d,u}[n]_\mathrm{opt}$, (\ref{D2UpathlossThetaSubOnlyTheta}) is non-increasing function to $t$ and $F(b) \le F(t) \le F(a)$.
If $\theta_{d,u}[n]_\mathrm{opt} \le a \le b$, (\ref{D2UpathlossThetaSubOnlyTheta}) is non-decreasing function to $t$ and $F(a) \le F(t) \le F(b)$.
If $a \le \theta_{d,u}[n]_\mathrm{opt} \le b$, (\ref{D2UpathlossThetaSubOnlyTheta}) ensures that $F(t) \le \max\{F(a), F(b)\}$.
Since that, we can argue that for any $a \in [0^{\circ},90^{\circ}]$ and $b \in [0^{\circ},90^{\circ}]$:
\begin{equation}
\begin{aligned}
F(\lambda{a} + (1-\lambda)b) \le \max\{F(a), F(b)\} \quad \forall \lambda \in [0,1]
\end{aligned}
\label{proofQuasiConvex}
\end{equation}
which corresponds to the definition of quasi-convex function.
Therefore, (\ref{D2UpathlossThetaSubR}) is a quasi-convex function with only one global minimum.
\label{appendixA}
\end{appendices}

\bibliography{referenceSWS}
\bibliographystyle{IEEEtran}

\end{document}